\documentclass[10pt, journal]{IEEEtran}
\def\ps@headings{%
\def\@oddhead{\mbox{}\scriptsize\rightmark \hfil \thepage}%
\def\@evenhead{\scriptsize\thepage \hfil \leftmark\mbox{}}%
\def\@oddfoot{}%
\def\@evenfoot{}}
\makeatother
\usepackage{amssymb,amsmath,amsthm}
\usepackage{graphicx}
\usepackage{esvect}
\usepackage{tabularx}

\usepackage{amsmath}
\usepackage{changes}
 \newcommand{\rem}[1]{}

\scriptsize

\usepackage[TABBOTCAP]{subfigure}
\usepackage{mdwmath}
\usepackage{mdwtab}
\usepackage{soul}
\usepackage{cite}
 \usepackage{multirow}
 \usepackage{color}
  \usepackage[linesnumbered,ruled]{algorithm2e}
\usepackage{setspace}

\begin{document}

\title{Exploiting Parallelism in Optical Network Systems: A Case Study of Random Linear Network Coding (RLNC) in Ethernet-over-Optical Networks}
 \author{\IEEEauthorblockN{Anna Engelmann$^*$, Wolfgang Bziuk$^*$, Admela Jukan$^*$ and Muriel M\'{e}dard$^{**}$}\\
\IEEEauthorblockA{Technische Universit\"at Carolo-Wilhelmina zu Braunschweig, Germany$^*$\\
Massachusetts Institute of Technology, USA$^{**}$ }
}
\maketitle

\begin{abstract}
As parallelism becomes critically important in the semiconductor technology, high-performance computing, and cloud applications, parallel network systems will increasingly follow suit. Today, parallelism is an essential architectural feature of 40/100/400 Gigabit Ethernet standards, whereby high speed Ethernet systems are equipped with multiple parallel network interfaces. This creates new network topology abstractions and new technology requirements: instead of a single high capacity network link, multiple Ethernet end-points and interfaces need to be considered together with multiple links in form of discrete parallel paths. This new paradigm is enabling implementations of various new features to improve overall system performance. In this paper, we analyze the performance of parallel network systems with network coding. In particular, by using random LNC (RLNC), -- a code without the need for decoding, we can make use of the fact that we have codes that are both distributed (removing the need for coordination or optimization of resources) and composable (without the need to exchange code information), leading to a fully stateless operation. We propose a novel theoretical modeling framework, including derivation of the upper and lower bounds as well as an expected value of the differential delay of parallel paths, and the resulting queue size at the receiver. The results show a great promise of network system parallelism in combination with RLNC: with a proper set of design parameters, the differential delay and the buffer size at the Ethernet receiver can be reduced significantly, while the cross-layer design and routing can be greatly simplified.
\end{abstract}

\section{Introduction}

\par { The high-speed Ethernet standard IEEE802.3 specifies that the 40/100/400Gb/s Ethernet traffic can be packetized and distributed over multiple parallel lanes (e.g., 10 lanes x 10Gb/s, for 100Gb/s), also referred to as Multi-Lane Distribution (MLD) \cite{802.3ba}. Each lane can then be mapped onto parallel optical channels for transmission in Optical Transport Networks (OTN). Thus, each Ethernet end point is attached to network through multiple and parallel logical points of attachments, -- which can be dynamically configured, enabling a functional decomposition of the overall system and new topological abstractions where multiple end points need to be mapped to multiple paths in the networks. Also other network systems already support parallelism: network parallelism as a concept can extend from networking to both the commodity hardware and modern cloud computing, such as multi-core parallel architecture and Hadoop Map Reduce.   As the foundational capabilities for the parallel network system mature, they carry the potential to become the driving engine for multiple facets of information technology infrastructure today, including physical layer security, scalable data center architectures, and network load balancing. 

\par Parallel network systems are however more complex than corresponding systems with serial connections and single end-system interfaces. The complexity needs to be evaluated against the benefits of parallelization. Another issue is of performance, including packet skews and delays. The skew of data packets occurs due to diversity of parallel links, which in Ethernet receivers requires the so-called \emph{de-skewing} via buffering. As the standard IEEE802.3 defines the maximum of 180ns skew per Ethernet lane to eliminate or reduce retransmissions or dropping of the unrecoverable Ethernet frames, keeping the skew within bounds is critical. The challenge of delay and packet processing requires methods of reduction of data coding overhead in different ISO/OSI layers. Today, each layer currently has its own representation of coding, e.g., source coding or coded storage. A simple distributed coding over Ethernet layer and optical layer, for instance, can eliminate the delay and complexity caused by mapping different coding schemes for different purposes and at different layers. Hence, parallelism, coding and routing requires different thinking about the cross-layer system engineering.  

\par This paper sets the goal to explore the potential of parallelism in future network systems in combination with simple and unified coding in multiple layers, to solve system performance, cross-layer design and network resource utilization problem.  We choose random linear network coding (RLNC) as unified coding scheme between Ethernet and optical (physical) layers.  RLNC is known for its capability to allow a recoding without prior decoding resulting in two main advantages that eliminate the need i) for cross-layer coordination, or optimization of resources and ii) for exchange of code information. We take the following approach to system analysis and modeling.} A serial data traffic is split into discrete parallel parts (frames) in the electronic end-system (Ethernet) that are assigned several parallel optical interfaces and paths in the optical network. A distributed and composable cross-layer coordination is based on RLNC, which can be employed in both layers. In this system, we analytically derive the expected values of differential delay in a generic parallel network system, whereby the number of parallel paths in the network maybe equal or larger than the number of parallel Ethernet lanes in the end-system. We furthermore derive the upper and lower bounds on the resulting queue size at the receiver, with and without RLNC. {We analyze and compare the networks with optimal (without RLNC) and random routing (with RLNC) in the network, and show that RLNC can significantly reduce, and even eliminate, the path computation complexity and need for optimal routing. The theoretical results are validated by simulations, and show that the required storage at receiver in the parallel Ethernet system with RLNC is always smaller than in an Ethernet-over-optical system without RLNC, and that irrespectively of the routing scheme. We also show that by carefully selecting the level of parallelism in the Ethernet and optical network systems, the cross-layer interactions can be designed in a simple and practical fashion.}

\par The rest of the paper is organized as follows. Section II discusses prior art and summarizes our contribution. Section III presents the parallel network system model. Section IV focuses on modeling of the expected differential delay and derives lower and upper bounds on skew queues at the receiver. Section V shows analytical and simulation results. Conclusions are drawn in Section VI.

\section{Related work and our contribution}

\par { From the functional perspective, some aspects of previous work bear resemblance with our concept of multichannel
transmission and multi-interface nodes, but none of the previous work is directly applicable without a
critical consideration. Various signal multiplexing technologies, for example, Wavelength Division Multiplexing
(WDM), Polarization Division Multiplexing (PDM), Space Division Multiplexing (SDM) or elastic optical
Frequency Division Multiplexing (OFDM), differ in the ways to realize physical links and impact the system
design differently. Given the broad range or related topics, this section reviews those aspects of the state of the
art that we find relevant to the idea of parallelism in optical network systems, whereby specific parts of it, like
RLNC, are to be seen as tools used in combination with, and not as solutions for, network system parallelism.
} 


\subsection{Multilane Distribution in High-Speed Ethernet}
\par  The 100/400~GE standards in IEEE 802.3~\cite{802.3ba, 802.3bs} define Multiple Lane Distribution (MLD) systems with parallel interfaces. In these systems, high-speed Ethernet signals are distributed onto multiple lanes in a round robin fashion, with data rate at 40/100~Gbps perfectly compatible with optical channel rates in Optical Transport Networks (OTNs)~\cite{ITU-T:G.709, ITU-T:G.7715}. It should be noted that MLD in high-speed Ethernet defines cross-layer system requirements different from inverse multiplexing. The latter technique is standardized in IEEE802.3 as the so-called \emph{Link Aggregation,} supported by the  Link Aggregation Control Protocol ({LACP}). In optical transport networks (OTN) and synchronous optical network (SONET) standards, the inverse multiplexing is also defined, as Virtual Concatenation (VCAT).  In fact, MLD does not require inverse multiplexing techniques, albeit their proven ability to implement dynamic skew compensation mechanism as studied in \cite{Yang:2009} and \cite{Sun:2008}. Instead, MLD enables parallel lanes to be mapped to parallel interfaces and channels in the optical layer, allowing the implementation and management of cross-layer interactions, similar to what has been shown Layers 3 and above \cite{6692295}. Past work also used Layer 2 switching concepts, particularly OpenFlow, in conjunction with multipath TCP (MPTCP) \cite{van2012multipathing}.



\subsection{Parallelism vs. multipath routing}

\par Our approach utilizes concepts known from similar to multipath routing in layer 3, but extends the same to the network with multiple end-system interfaces. The number of the multiple network end-points can be dynamically configured, which creates not only new network abstractions, but also new routing optimization problems, since the number of end-points is usually matched to the number of routes in the network. In high-speed Ethernet systems, frames are distributed onto multiple lanes (4, or 10 lanes) in a round robin fashion, with data rates perfectly compatible with the corresponding number of optical channel rates in Optical Transport Networks (OTNs)~\cite{ITU-T:G.709, ITU-T:G.7715}. 

\par In the optical layer, the \emph{elastic (cognitive, flexible) optical networks}, optical transponders (OTP) have evolved from fixed or mixed-line-rate (MLR) to bandwidth-variable (BV) to sliceable-BV \cite{Jinno2012,tanaka2014recent} to serve various low-capacity (e.g., 40G) and high capacity (400G) demands. The \textit{sliceable} bandwidth-variable transponders (SBVTs), also called multiflow optical transponder (MF-OTP), maps traffic flows coming from the upper layers (e.g., Internet flows) to multiple optical flows. Moreover, recent work showed that parameters of multiflow optical transponders can be software programmed \cite{sambo2014programmable}. These systems are different from parallel network systems, since the Internet traffic flows cannot be dynamically configured to map  the optical routes, for instance with an software defined routers. 

\par It should be noted that prior art proposed optimizations to compute multiple paths. In contrast, in our approach abandons the idea of routing optimizations,  for the reasons of complexity and also because current multipath routing algorithms cannot be used in network topologies with multiple links between nodes. The latter requires different approach to path analysis, and in our approach we use a combinatorial path analysis for the same. We also show that parallelism can simplify or eliminate routing when used in combination with random linear network coding, which is a significant result.

\subsection{Random Linear Network Coding (RLNC)}

\par {Previous work on linear network coding focused in general on  improving network throughput and reliability. However, significant body of work in the last decade (e.g., \cite{Eryilmaz:2006, Lucani:2012, Sorour:2010, Wang:2017, Chen:2016}) addressed with network coding the end-to-end delays improvement in delay-constrained networks in broadcast and unicast scenarios. In \cite{Eryilmaz:2006}, for instance, the delay performance of network coding was studied and compared it to scheduling methods. Lucani et. al, in \cite{Lucani:2012} tailored coding and feedback to reduce the expected delay. Paper \cite{Sorour:2010} studied the problem of minimizing the mean completion delay for instantly decodable network coding. In \cite{Wang:2017, Chen:2016} authors showed that network coding can outperform optimal routing in single unicast setting. More recent works, like \cite{Cloud:2016}, presented a streaming code that uses forward error correction to reduce in-order delivery delay over multiple parallel wireless networks. However, none of these works address delay in parallel network systems. } 

\par Network-coded multipath routing has been applied for erasure correction~\cite{Maxemchuk:2007}, where the combined information from multiple paths is transferred on a few additional (parallel) paths. The additional information was used to recover the missing information during decoding. 


\par In optical networks, our previous work~\cite{6831430} proposed for the first time a \emph{network coded parallel transmission} scheme for high-speed Ethernet using multipath routing. Paper \cite{Chen:2014}  focused on enabling  parallel transmission by  linear network coding without consideration of data link layer technology. In \cite{ICC2015} we presented a preliminary theoretical model  to achieve fault tolerance by using 2-parallel transmission and RLNC to achieve better spectral efficiency in the optical layer. Finally, in \cite{Engelmann:2016}, we showed that utilizing of RLNC significantly improve reliability and security in parallel optical transmission systems. 

{Our cross-layer approach can be generally based on any symbol based  MDS-Codes (Maximum Distance Separable Code), while we decided to use random LNC as a tool owing to the fact that it allows decoupling between code selection and transmission architecture. RLNC encoding and decoding can be perform in a parallel fashion \cite{Vingelmann2010}, whereas strucutured MDS codes are generally difficult to code or decode in a multithreaded fashion. The distributed nature of the RLNC code construction removes the need for cross-layer coordination when combined with parallelization. Different parallel channels may construct their own codes, and further parallelization takes place with the use of a single coding approach, without the need for state awareness across parallel paths. Our choice of RLNC moreover was motivated by potential use of its recoding capability and design of unified code for a network system in a cross-layer fashion. The composability feature underlies the ability to have cross-layer operation without the need to exchange state or other code-related information across layers. Each layer may, or not, introduce its own coding, composing upon the coding of other layers. Even after repeated coding (recoding), the effect upon the data is of a single linear transformation, requiring no decoding in the network, but a single linear inversion. This makes the combination of RLNC and parallelism especially promising.  Note also that RLNC may lend itself to a hybrid use where some of the data may be transmitted uncoded. We do not present that scenario explicitly.}

\subsection{Our contribution}

\par This paper builds on our preliminary works \cite{6831430, Chen:2014}. {In extension of the preliminary work, this paper provides
\begin{itemize}
\item Derivation of the expected value of the differential delay in arbitrary networks, and between any pair of arbitrary nodes connected with multiple parallel links, enabling routingless (or, random routed) network operation, in cases 1) without coding, 2) with RLNC and 3) with coding redundancy.
\item Derivation of occurrence probability of maximum possible differential delay, including cases where network contains multiple links and paths with maximal or/and minimal possible delay, which is a case study of practical relevance;
\item A new theoretical framework to queue analysis in end-systems including the derivation of a closed form of expected buffer size at receiver, with and without RLNC, and for an arbitrary distribution of path delays;
\item Analysis of the impact of coding redundancy and the level of parallelism on the network performance, and buffer sizing at receiver;  
\end{itemize}
}

\section{System Model}\label{algo}

\begin{figure*}[!th]
\vspace{-0.5cm}
\includegraphics[width=2\columnwidth]{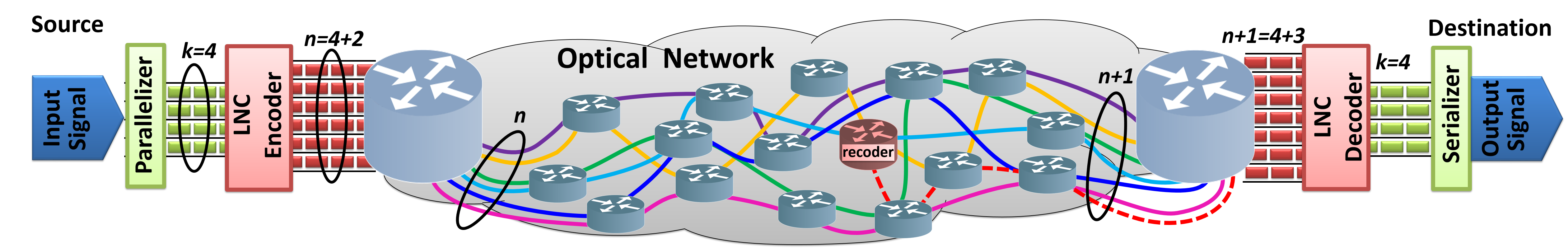}
\vspace{-0.1cm}
\caption{{A parallel network system architecture. ($k$ is the number of parallel flows (sub-flows, or lanes), $n$ is the number of utilized optical paths in the network; each pair of nodes is connected with 10 parallel interfaces and links in the network.)}}
\label{network}
\end{figure*}

\subsection{Background}
\begin{figure*}[!th]
\subfigure[]{
\includegraphics[width=1\columnwidth]{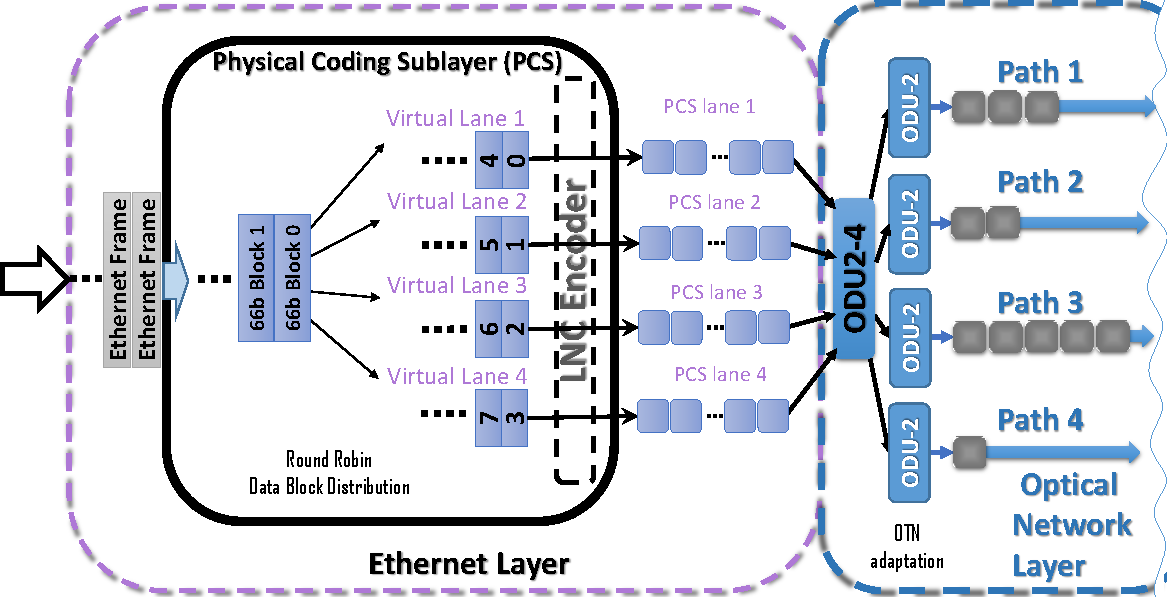}
\label{transmitter}
}
\subfigure[]{
\includegraphics[width=1\columnwidth]{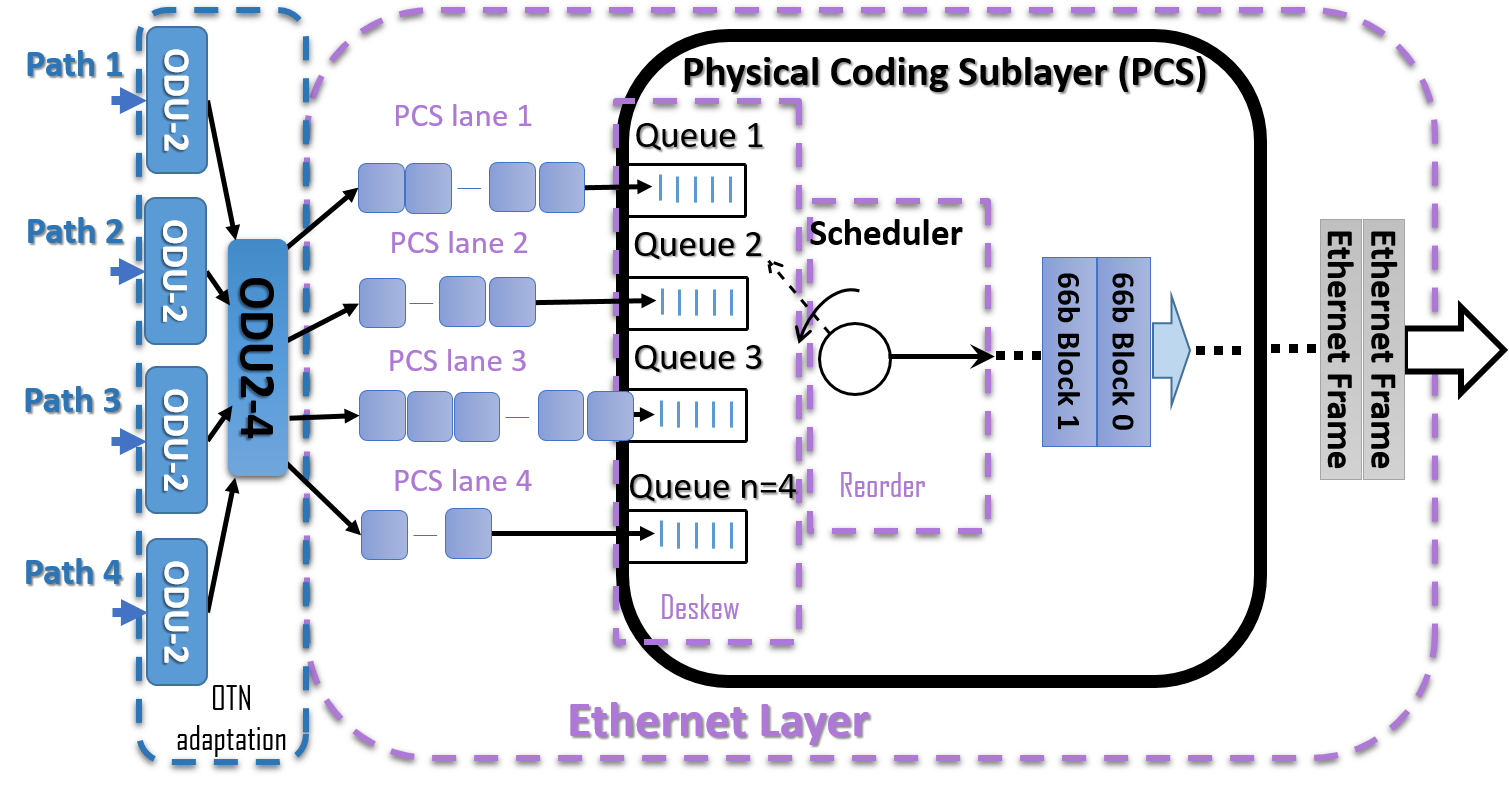}
\label{receiver}
}
\vspace{-0.35cm}
\caption{Multi-lane Ethernet-over-optical network system. (a) Traffic distribution in source; (b) Deskew buffer from \cite{802.3ba}.}\label{buffers}
\vspace{-0.4cm}
\label{system}
\end{figure*}

{Fig.~\ref{network} illustrates a parallel network system architecture envisioned. At the sender, a serial flow of high-speed data units is split into up to $k$ parallel flows (sub-flows, or lanes), whereby each parallel flow is then independently routed over $n$ separate optical channels. Depending on the application, the number of electronic lanes can be perfectly matched to the number of optical channels ($k=n$), whereby the aggregated capacity in the optical layer is always greater or equal to those in the electronic layer. Fig.~\ref{network} also illustrates how special functions can be added to each parallel channel, such as adaptive coding
features. The input data units (e.g., packets, frames)  are encoded using an RLNC encoder. The number of parallel lanes, $k$, and RLNC are related as follows: data units from $k$ parallel flows encoded with the same set of coding coefficients are called a \emph{generation,} while the number of resulting parallel flows after decoding $n \geq k$ is defined as \emph{generation size}, here $k=4, n=6$. We refer to the number of resulting parallel lanes, i.e., $n$, as the \emph{level of parallelism} in end-system ($n=6$). The network topology is such that every node is connected with other nodes over mulitple parallel links (here: 10 parallel links between any pair of nodes). The number of encoded data units $n=6$ is generally equal to the number of parallel paths and interfaces allocated in the network, whereby in our example the source node uses 6 out of 10 parallel interfaces to setup 6 parallel paths. If $n>k$, as we illustrate here, we refer to $r = n-k$ as \emph{redundancy}, here $r=2$. The decoder starts the decoding as soon as at least $k$ data units from one generation arrived. We envision that forwarding nodes in the middle of network can perform additional recoding of optical flows to the same destination as shown in Fig.~\ref{network} (dashed line). With recoding, it is possible to insert additional redundancy and so increase fault tolerance, without decoding.} 

\par Fig.~\ref{system} shows a typical multi-lane 40Gb/s Ethernet-over-optical network system\cite{802.3ba}. In the transmitter, a high speed stream of serial Ethernet frames
{is split into data blocks of 64b, encoded with 64b/66b line code in Physical Coding Sublayer (PCS), and distributed over $k=4$  virtual Ethernet lanes. For identification of the lane ordering at the receiver, specific alignment markers are inserted in each lane after $M_L=16383$ data blocks. After that, each 10GE Ethernet lane is mapped to four optical data units (ODU), here of type 2. The ODU2e method enables the transparent 10GE mapping using an over-clocking approach, whereas extended Generic Framing Procedure (GFP) principles are applied. The ODU signals are then modulated on four optical carriers and transmitted over four optical channels (Path 1, 2, 3, 4). In general, the number of Ethernet virtual lanes and the allocated optical paths do not need to be equal. However, in our model, the OTN concept is assumed to generally map data streams from Ethernet lanes into $n$ ODU2-$n$v or ODU2e-$n$v containers.}

\par A simplified architecture of the receiver, also according to IEEE802.3ba, is shown in Fig. ~\ref{receiver}. Here, PCS layer processes the  66b data blocks received to retrieve the original Ethernet frame, which requires multiple processing entities, including lane block synchronization, lane deskew and reorder, alignment removal, etc. (not all shown here). To illustrate this, let us  assume that paths $3$ and $4$ are the shortest and the longest path in terms of delay, respectively. For compensation of the resulting inter-lane skew, the data blocks from path $3$ must be buffered in the receiver until data from longer paths arrive, i.e. paths 1, 2, and 4. For compensation, the receiver implements the so-called \emph{deskewing} and \emph{reordering}. The deskew function of the PCS is implemented with the input FIFO queues, {that store data blocks until the alignment markers of all lanes are received and synchronized. This allows the scheduler to start the lane identification, alignment removal and the reordering to form the original data stream.}

\par
\begin{figure} [!t]
 \centerline{\includegraphics[width=1\columnwidth]{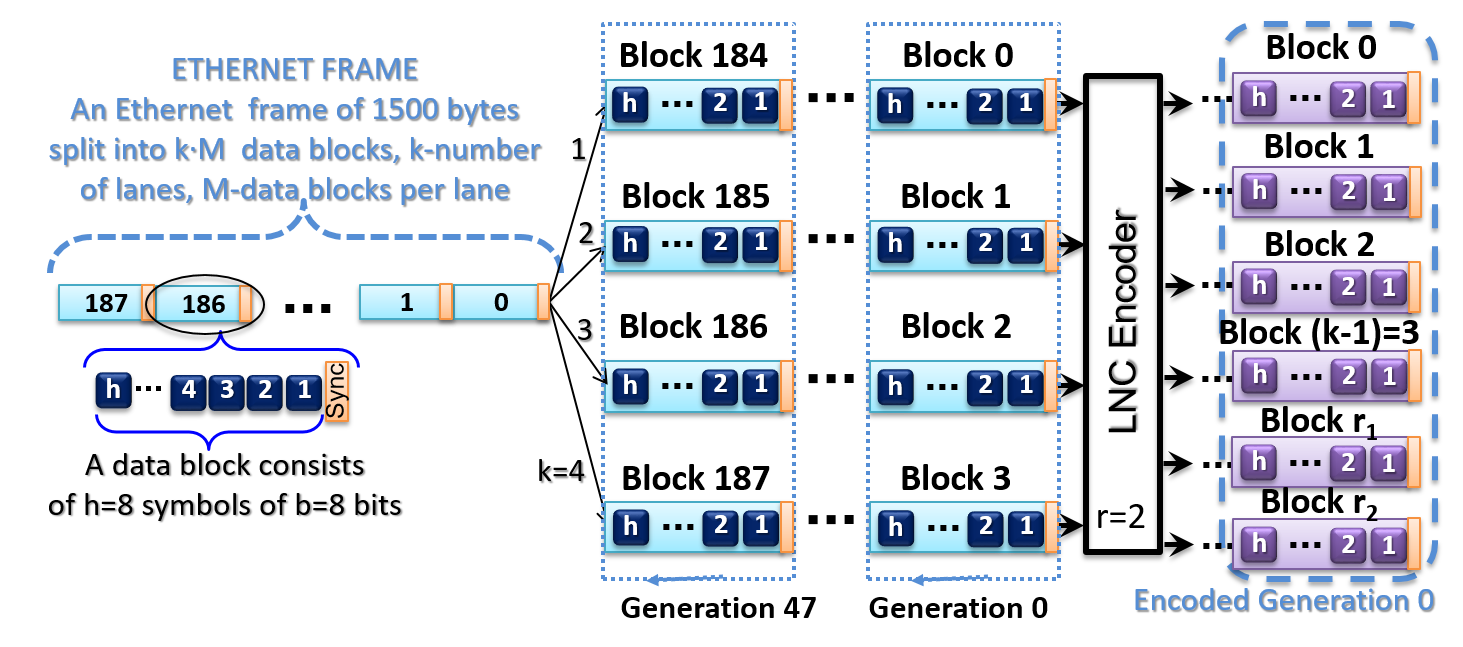}}
 \caption{Ethernet traffic parallelization with LNC.}
\label{traffic}\vspace{-0.4cm}
 \end{figure}

 Let us now focus on the receiver design with RLNC. A nice feature of RLNC is that it can be implemented without altering the system architecture presented, see dashed box in Fig.~\ref{transmitter}. 
{The coding process is illustrated in Fig.~\ref{traffic} in more detail. Let us assume an Ethernet frame of  $12000$ bits ($1500bytes$) split into $188$ data blocks and then encoded with 64b/66b line code  \footnote{The line code is not to be mistaken for RLNC. The latter is performed over the last 64b after sync header.}, where we introduce the notation 64b/64b+2b to differentiate between 64 data bits and the 2 bits of the sync header according to \cite{802.3ba}. The data blocks are then distributed over  $k=4$ virtual PCS lanes so that each sub-flow on defined virtual (parallel) lane contains exactly $47$ data blocks\footnote{In our model of traffic splitting, we assume the bit padding in that case that the data block, or symbol, is incomplete.}.}

\begin{figure} [t]
 \centerline{\includegraphics[width=1.0\columnwidth]{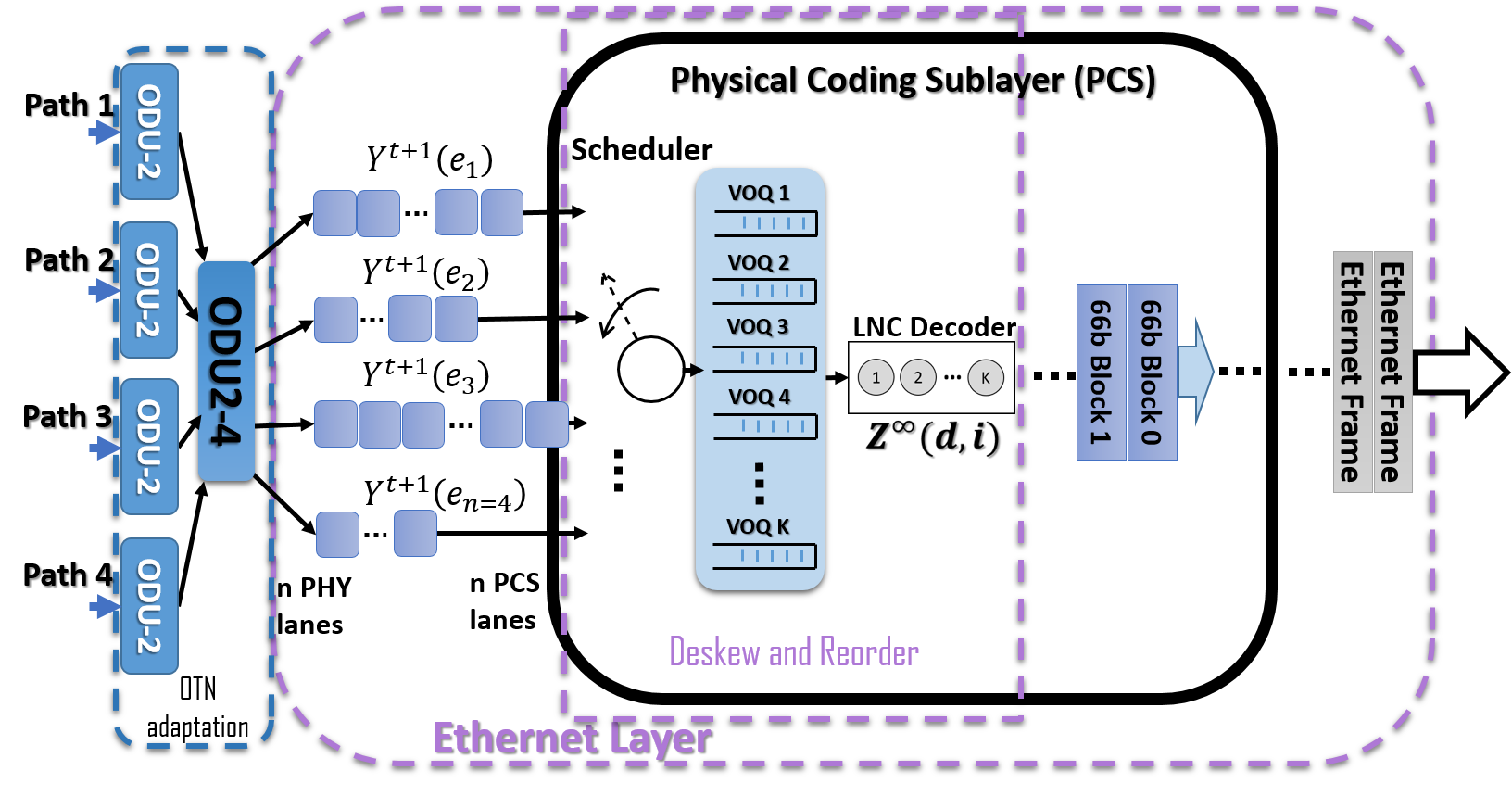}}
 \caption{Decoding buffer at the receiver.}
\label{DecBuffer}\vspace{-0.5cm}
 \end{figure}

{Assuming the RLNC coding process is based on symbol size $b=8$ bits, and since blocks contain $64$ data bits excluding the 2 synchronization bits, each coded data block would contain $h=8$ symbols. All $k$ symbols from each parallel lane related to parallel data blocks that are simultaneously encoded with the same set of RLNC coefficients (a \emph{generation}), while the number of resulting encoded data blocks, i.e., $n$, is defined as \emph{generation size}. In Fig.~\ref{traffic}, each Ethernet frame thus encoded into $47$ generations, while the generation is extended by two redundant blocks resulting in generation size $n=6$. The 2 sync header bits of each data block bypass the RLNC encoder and are added after coding as header in the form 64b/64b+2b+Cb, where C is an additional ID-header.

\par {At the receiver, the reference deskew and reorder model with RLNC is shown in Fig.~\ref{DecBuffer}, where transmission is over $n=k=4$ paths similar to traditional Ethernet system (Fig.~\ref{receiver}), i.e., without redundancy. Later (Fig. \ref{decodbuffer}), we discuss the system implemented with coding redundancy.  {The distributed line buffer of the Ethernet system is now organized as a centralized decoding buffer consisting} of multiple virtual output queues (VOQ), whereby a new VOQ is created for each generation, each time the first data block of a new generation arrives. The decoder checks the existing VOQ for complete generations, and starts decoding as soon as one generation is complete, whereby all data blocks of a complete generation are decoded in parallel{, by running Gaussian elimination. Thus the parallel decoding replaces the line specific deskewing approach of the Ethernet system by taking advantage of the multiplexing gain due to the centralized buffer}. After decoding, the data blocks are sent in the correct order, -- thus eliminating the need for reordering. That is due to the fact that data blocks are decoded in parallel, while a correct assignment of decoding coefficients to encoded data blocks assures the right order of data blocks. As a result, decoded data blocks are only  serialized.

\par For successful decoding, all data blocks from the same generation $g_\nu$ need to be uniquely identified $b_{j_\nu}, j=1,...,n$. This can be implemented using additional $C=6$ bits in the header to form 72b coded blocks. Each data block of a generation $g_\nu$ is identified by the same number $b_{j_\nu}=l, j=1,...,n$, whereby we use $L=2^C$. At the receiver, the identifier is processed by the scheduler and addresses the correct VOQ $l$. Since ODU2 payload includes 15232 bytes, after $L=64$ sequentially arrived data blocks per lane, the number wrap-round will overwrite the same VOQ with a new generation. This corresponds to the maximium delay difference between lanes to $64 tu$, where the time unit (tu) is the transmission time of a data block. For instance, for 10Gbit/s we have $tu=7.2$ns and a maximum delay difference of $446.2$ns, which is far larger than the required $180$ns for Ethernet systems, and thus would be unacceptable. \footnote{Note, the 72b=64b/64b+2b+6b block increases the line rate by $12/11$, which requires to develop an efficient GFP method in the OTN layer.} A sensible approach to address this is to reuse the Ethernet inherent alignment marker process without adding additional ID-header bits, i.e. $C=0$, also in line with the existing standards. After receiving the first alignment marker, the corresponding lane is marked as reference, the scheduler initialized the VOQ $1$ for the first generation and for each following block of the same line the next VOQ $l$ is generated. In fact this allows to address $M_L=16383$ different VOQs, and to compensate delay differences of up to $16383 tu$. To limit the buffer space, however, similar to the method with ID-header, we could cyclically overwrite the VOQs after receiving the $L^{th}$ data blocks on the initializing lane, where $L<<M_L$. \footnote{The data blocks extracted from an ODU container or its sync header may be erroneous, despite the existence of FEC in the optical layer. In case of a single block errors, the RLNC decoding of one generation will fail. If an alignment marker is erroneous, all data blocks received on the associated lane may be sorted to the wrong VOQ resulting in decoding errors. This error process will be stopped with arrival of the next marker and, thus, re-initialization of VOQs mapping. At the same time, this issue is not different from the erroneous packet handling in the conventional Ethernet system.}}


\par  The coding coefficients required for encoding and decoding can be selected and distributed by control plane using an out-of-band signaling channel. An in-band signaling method can be used by applying transcoding, as specified in IEEE802.bj. For example with 4 PCS lanes, the 256b/257b transcoding enables us to transmit 35 additional bits after sending 5 blocks of 256b/257b (or after 20 blocks of 64b/66b) per lane (in total, 140 additional bits serially) without increasing the 10GE line rate, which is inline with the Ethernet and OTN standards. Thus, $k=4$ coding vectors each of length 8b can be sent on each lane every 20 64b/66b blocks to the destination, i.e., 20 successive generations are coded with the same coding vector. 
}
\par In the model that follows, encoding and decoding are applied in the end-systems, i.e., on the Ethernet layer, while optical nodes in the core network simply forward the incoming coded data over reserved outgoing interfaces. 

%
%
%

\subsection {RLNC-based end-system model}

\par To model the RLNC based  end-to-end system, we adopt the network model from \cite{Koetter_Medard} representing  a network as a directed and acyclic graph $G (V, E)$, where  $V$ and $E$ are vertex set and edge set, respectively. The source and destination nodes are denoted as $s\in V$ and $d \in V$, respectively.  A distinction is made between incoming and outgoing links of an arbitrary node $v\in V$, which are denoted as a set $\mathcal E_{\textrm{in}} (v)$ and  $\mathcal E_{\textrm{out}}(v)$, respectively. {A link $e_i$ is an incoming link $e_i\in \mathcal E_{\textrm{in}} (v)$ of a node $v$, if $head (e_i)= v$, where $1\leq i\leq |\mathcal E_{\textrm{in}} (v)|$, while link $e_j$ is an outgoing link $e_j \in \mathcal E_{\textrm{out}} (v)$ of a node $v$, if $tail(e_j)=v$, where $1\leq j\leq |\mathcal E_{\textrm{out}} (v)|$.}


\par As illustrated in Fig.~\ref{traffic}, the traffic sequence, is decomposed into \emph{data blocks} of the same length, i.e., $h$ symbols each. The linear coding process is performed over a field $F_{2^b}$, whereby each symbol has the same length of $b$ bits. 

\par We define \emph{time unit (tu)} as a discrete time based on the link capacity of the physical link, which can be analyzed as a transmission delay of one data block. Thus, the parallelization, reordering and de-skewing at the end systems can be modeled as a discrete time process. At time $t$, the incoming symbols $\nu$, $\nu=1,2,...,h$, of parallel data blocks $\vv x_i$ at source node $s\in V$ are generated by the processes $X^t(s, i)=x_{i\nu}$, on every virtual lane $i$, denoted as $e_i$, $1\le i\le k$, 
$head (e_i)= s$.  The incoming symbols $x_{i\nu}$ are encoded into symbols $y_{j\nu}$, $1\le j\le n$, and sent out on each outgoing lane $e_j \in E$, $tail(e_j)=s$. In this model, the RLNC encoder buffers incoming symbols from all $k$ lanes in parallel, and encodes the same with simple linear coding. Thus, the signal carried on an outgoing link $e_j$ of source $s$ at time $t+t_{\delta}'$ is:
    \begin{equation}\label{source}
    \forall e_j: tail(e_j)=s:   \quad Y^{t+t_{\delta}'}(s,j) = \sum_{i=1}^{k} a_{ij} \cdot X^{t}(s, i),
\end{equation}
where $t_{\delta}'$ is an encoding interval and $a_{ij}$ collected in the matrix $\boldsymbol{A}$ are encoding coefficients from the finite field $F_{2^b}$.

\par  At the receiver, the decoded information at time $t+t_{\delta}$ on parallel lane $i$ is modeled as:
\begin{equation}\label{decodingDest}
Z^{t+t_{\delta}}(d, i) = \sum_{e_j: head(e_j)=d} b_{i, j } \cdot Y^{t}(d,j),
\end{equation}
where $t_{\delta}$ is decoding interval, $b_{i, j}$ are coding coefficients from the finite field $F_{2^b}$ collected in the matrix $\boldsymbol{B}$ \cite{Koetter_Medard}. {Generally, RLNC can result in linearly dependent combinations, whereby the probability for that combinations is related to the finite field size.  The probability of selecting coefficients that do not correspond to a decodable code instance is of the order of $2^{-8}$ \cite{TBF11, ho_random_2006, CVC13, Li2015, HPW13, Lee2014}. Thus, with high probability, regardless of the number of paths selected, RLNC will lead to a satisfied LNC. The decobability can be verified at the transmitter or receiver. In case an instance of coding coefficent is not decodable, the encoder can readily select another instance, with coefficients selected uniformly at random from the field in which we operate. }



\subsection {Network model}

\par The network $G(V,E)$ is modeled as a directed graph with a set of $V$ nodes and $E$ links, whereby each pair of nodes is connected with multiple parallel links. In this topology, it is expected that at least $N\geq n\geq k$ out of $F$ existing parallel links and paths are available between source $s$ and destination $d$, whereby only $n\geq k$ paths are selected for parallel transmission. To derive the likelihood that network can provide $N\geq n$ paths, we use the following model applicable to connection-oriented networks. {A network $G(V,E)$ can provide at most $N\leq min\{|\mathcal E_{\textrm{out}} (s)|, |\mathcal E_{\textrm{in}} (d)|\}$ available parallel paths between nodes $s$ and $d$}, while the setup probability of an arbitrary path $\mathcal P(s,d)$ over a defined link denoted as $P_{\textrm{setup}}$\footnote{Since our motivation is to analyze the differential delay in the network, and the resulting buffer size at the receiver, we do not consider in this paper how various traffic load pattern impact the path setup probability $P_{\textrm{setup}}$.}. We approximate the model for blocking probability of the connection request $B(n)$ by assuming that the network load is distributed so that each out of $F$ possible paths can be setup with an equal probability, $P_{\textrm{setup}}$. As a result, the probability, that a path $\mathcal P(s,d)$ between $s$ and $d$ cannot be setup, and is blocked, is defined as
\begin{equation}\label{blocking}
 P_{\textrm{B}}=1-P_{\textrm{setup}}
 \end{equation}

\par Thus, the probability for $N$ available parallel paths out of $F$ existing paths would follow the Binomial distribution, i.e.,
 \begin{equation}\label{ProbNnew}
Pr(N=j)=\binom{F}{j}(1-P_{\textrm{B}})^j P^{F-j}_{\textrm{B}}
\end{equation}
Finally, the transmission request is blocked with probability $B(n)$, when the number of available paths $N$ is lower than the number of outgoing interfaces $n$, i.e., $N<n$.
 \begin{equation}\label{ProbBlockedTrans}
B(n)=\sum_{j=0}^{n-1}Pr(N=j)
\end{equation}
The mean number of available paths is determined as
 \begin{equation}\label{meanPaths0}
\bar {N} =\sum_{j=0}^{F}j\cdot Pr(N=j)=F\cdot P_{\textrm{setup}}
\end{equation}
On the other hand, the mean number of available for optimization or random selection paths $E\{N|N\geq n\}$, which ensures the successful parallel transmission and is relevant for the buffer analysis in the next Section, can be derived as

 \begin{equation}\label{meanPaths1}
E\{N|N\!\!\geq\!\! n\}\!\! =\!\! \sum_{j=0}^{F}\! j\cdot Pr(N\!\!=\!\!j|N\!\!\geq\!\! n)\!\!=\!\!\frac{\sum_{j=n}^{F}j\!\!\cdot\!\! Pr(N\!\!=\!\!j)}{1-B(n)}
\end{equation}

\par As previously mentioned, RLNC can extend the generation size by including $r$ redundant data blocks resulting in $r$ redundant data flows from source. {The redundant, i.e., $r$, data blocks are transmitted in parallel with other $k$ data blocks from the same generation}. In case of data block loss or network failures, a data block coming from redundant parallel paths can replace any data block from the same generation. We show that this feature is useful not only for fault tolerance but also for reducing the expected value of the differential delay, and thus the buffer size.

\section{Analysis}

The analysis includes three parts: (i) analysis of the expected differential delay in a generic network, (ii) analysis of the impact of coding redundancy on differential delay, and, (iii) derivation of the expected value of the queue (buffer) size at the receiver, including the upper and lower bounds.

{For presented analysis, we utilize following underlying assumptions
\begin{itemize}
\item The network does not exhibit any failures or losses;
\item In transmission system with RLNC, a set of $n$ parallel paths is chosen randomly at the source, and with the same probability among all $N\leq F$ paths available;
\item {Let us assume that $N$ available parallel paths $\mathcal P_l$, $l = 1, ..., N$ between source $s$ and destination $d$ in $G$ are collected in a set $G_{\mathcal P} (s,d) =  \{\mathcal P_{1},\mathcal P_{2}, ..., \mathcal P_{l},... , \mathcal P_{N}\} $ and are sorted in the ascending order so that the increasing index $l$ of each path $\mathcal P_{l}$  corresponds to an increasing path delay $d_{l}$, i.e., $d_{1} \leq  d_2   ... \leq   d_{l},..., \leq   d_{N}$, which are arranged in a vector of length $N$, $\vv{d}=( d_1, d_2, ..., d_{N})$. { Since we consider next only one certain source destination pair, the notation  $G_{\mathcal P} (s,d)$ can be simplified as $G_{\mathcal P} $};}  
\item Since any fiber path provides multiple wavelength path, we assume that multiple available paths can have the same end-to-end delay;
\item For a fair comparison, all data blocks in the system without network coding have the same size as in case with RLNC;
\item Link capacity is defined as a data block per \emph{time unit (tu)}. 
\item To simplify the analysis, we assume integer values for delays, i.e. $ \lfloor d_{i} \rfloor = d_{i}$, which can be realized without loss of accuracy by choosing a sufficiently small value of \emph{tu};  
\item We assume an idealized scheduler for both architectures, i.e.,  with and without RLNC; 
\item For scheduler, we do not assume a specific polling strategy, which may have an impact on the queue size.
\item In steady state, we assume the full traffic load and a deterministically distributed arrival process, where on each lane one data block arrives per \emph{tu}.
\item{We note the binomial coefficient $\binom{i}{j}$ as $C_{i,j}$, whereby $C_{i,j}=0$ for $j>i$.}
\end{itemize}
}

\subsection{Expected value of differential delay}

\par The differential delay $\tau$ is typically defined as difference in delays between the longest and the shortest paths $d_{\textrm{max}}$ and $d_{\textrm{min}}$, respectively. We next derive the expected value of differential delay given a number of available paths $N \leq F$ in arbitrary networks, whereby a set of $n$ parallel paths is chosen randomly. 


{ 
Let's assume that $n$ parallel paths chosen randomly with the same probability among $N$ available paths form a subset and let's denote $\mathcal M$ as the set of all  possible subsets. There are $|\mathcal M|=a_{\mathcal M}=C_{N,n}$  possible path combinations, where each combination is collected in a subset $\mathcal M_{n}(\alpha) \in \mathcal M$, $1 \leq \alpha \leq  a_{\mathcal M}$ and appears with same probability $P(\mathcal M_{n}(\alpha))= 1/a_{\mathcal M}$. However, all paths  in each subset  $\mathcal M_{n}(\alpha) $ are sorted so that their corresponding path delays appear in ascended order. This requires to derive an index mapping $l_m$, which maps an index $m$, $ m = 1,2,3,...,n$, used to specify a path $\mathcal P_{l_m}(\alpha)$ out of the subset $ \mathcal M_{n}(\alpha)=\{\mathcal P_{l_1}(\alpha),...,\mathcal P_{l_m}(\alpha),...,\mathcal P_{l_{n}}(\alpha)\}$ to a path  $\mathcal P_l \in G_{\mathcal P} $, $l=1,2,...,N$. This also maps the delay $d_{l_m}(\alpha)$ of  path  $\mathcal P_{l_m}(\alpha)$   to the corresponding component $d_l $ of the delay vector $\vv{d}$. This ensures, that the increasing index $m$ of each path $\mathcal P_{l_m}(\alpha)$ corresponds to an increasing path delay $d_{l_m}(\alpha)$, $m = 1,2,...,n$. 

To define such mapping let introduce $\mathcal N = \{1,2,... , N  \}$  and  $\mathcal N_{n} = \{ \Lambda_n \subset \mathcal N  : | \Lambda_n | = n   \} $, i.e. the set of all subsets of $\mathcal N$ with cardinality $n$. Let us use $\Lambda_n(\alpha)$, $\alpha = 1,2,...,a_{\mathcal M}$, to index each of these subsets. Thus based on paths $P_l \in  G_\mathcal P$, the set $ \mathcal M_{n}(\alpha)$ is given as  $\mathcal M_{n}(\alpha) = \{  \mathcal P_l : l \in \Lambda_{n}(\alpha)  \}$ and the final mapping between path $\mathcal P_{l_m}(\alpha) \in \mathcal M_{n}(\alpha)$  and $\mathcal P_l \in G_{\mathcal P} $  is defined by the index function
\begin{equation}\label{map}
\delta_{\alpha}(l,m)\!\! = \!\!
\begin{cases}
1 & \text{if the } m^{th} \text{ element of }  \Lambda_n(\alpha)   \text{ is  }  l \\
0 & \text{otherwise}
\end{cases}
\end{equation}
where $l \in\mathcal N $ and $m = 1,2,...,n$.
}
{ Furthermore, Eq.(\ref{map}) enables the mapping of paths by the relation $\mathcal P_{l_m}(\alpha) \equiv \sum_{i} \mathcal P_{i} \delta_{\alpha}(i,m) $ and has the property $\sum_l \delta_{\alpha}(l,m) = 1$. 
}
This ensures that each path $\mathcal P_{l} \in G_\mathcal P $ can occur once in the path set $\mathcal M_{n}(\alpha)$.  For example, assume $n=3$ parallel paths are randomly chosen from the set $G_\mathcal P $, i.e., $\mathcal P_{1},\mathcal P_{3}$ and $\mathcal P_{4}$. The sorting due to increasing delays defines the mapping to the selected subset shown by the equivalence  $\mathcal M_{n}(\alpha)=\{\mathcal P_{1_1}(\alpha),\mathcal P_{3_2}(\alpha),\mathcal P_{4_{3}}(\alpha)\} \equiv \{\mathcal P_{1},\mathcal P_{3},\mathcal P_{4}\}$.  

\par The longest and the shortest paths, $d_{l_{n}}(\alpha)= d_{\textrm{max}}(\alpha)$ and $d_{l_1}(\alpha)= d_{\textrm{min}}(\alpha)$, respectively, within chosen set $\mathcal M_{n}(\alpha)$  define the differential delay of the path set chosen, i.e.,
\begin{equation}\label{tau}
  \tau(\alpha) = d_{\textrm{max}}(\alpha)-d_{\textrm{min}}(\alpha)
\end{equation}
As we consider networks with a large number of path sets available, the differential delays depend on the path set chosen, i.e.,  $\alpha$, $\alpha = 1,...,a_{\mathcal M}$. The expected value of differential delay $\bar\tau=E_\mathcal M \{\tau(\alpha)\}=E_\mathcal M \{d_{\textrm{max}}(\alpha)-d_{\textrm{min}}(\alpha)\}$ takes into account all possible paths combinations, as indexed by $\mathcal M$, i.e., 
\begin{equation}\label{taumean}
 \bar\tau= E_{\mathcal M}\{\tau(\alpha)\} = E\{d_{\textrm{max}}(\alpha)\}-E\{d_{\textrm{min}}(\alpha)\}
\end{equation}
where we use $E_\mathcal M \{\cdot\} =E\{\cdot\}$ to simplify notation. The expected value $\bar\tau$ can be derived knowing the expected delay $E\{d_{m}(\alpha)\}$ of the $m^{th}$  path $\mathcal P_{l_m}(\alpha)$ over all path sets $\mathcal M_{n}(\alpha)$ randomly chosen with probability $P(\mathcal M_{n}(\alpha))= 1/a_{\mathcal M}$. Using the mapping  
{ $d_{j_m}(\alpha)=\sum_{l}  d_l \delta_{\alpha}(l,m) $ 
} 
as specified by Eq.~\eqref{map}, the expected delay of $m^{th}$ path   can be derived from

{
\begin{equation}\label{pmmean}
E\{d_{m}(\alpha)\}\!\!= \!\! \! \sum_{\alpha} \sum_l \!\!  d_l \delta_{\alpha}(l,m) P(\mathcal M_{n}(\alpha)) \!\!= \!\! \sum_l \!\! d_l p_l(m)
\end{equation}
where $p_l(m)=  1/a_{\mathcal M}    \sum_{\alpha} \delta_{\alpha}(l,m)$ 
} is the probability, that the $l^{th}$ path $\mathcal P_l$ with delay $d_l$ from $\vv d$ is selected as $m^{th}$ path with delay $d_{l_m}(\alpha)$, $m=1,2,...,n$, in a random chosen path set $\mathcal M_{n}(\alpha)\in \mathcal M$. {Using combinatorial  theory, in the range $ m\le l \le N-n+m$ this probability is given by 
\begin{equation}\label{ProbTi}
    p_l(m) = \frac{1}{a_{\mathcal M}} C_{l-1,m-1}C_{N-l,n-m}  
\end{equation}
The detailed derivation of Eq.(\ref{ProbTi}) is in Appendix. }Using Eqs. \eqref{pmmean}-\eqref{ProbTi}, the expected value of path delay of the $m^{th}$ path of a randomly chosen path set $\mathcal M_{n}(\alpha) \in \mathcal M$ yields,
\begin{equation}\label{ExpectedTi}
  \begin{split}
E\{d_{m} & (\alpha)\} =  \sum_{l=m}^{N-n+m} d_{l} \cdot p_l(m) \\
          &=  \frac{1}{a_{\mathcal M}}\sum_{l=m}^{N-n+m} d_{l} C_{l-1,m-1}C_{N-l,n-m} \\
          &= \frac{1}{a_{\mathcal M}}\sum_{l=0}^{N-n} \!\!   d_{(l+m)} C_{l+m-1,m-1}C_{N-l-m,n-m},
  \end{split}
\end{equation}
where $1\leq m\leq n$.
Thus, the expected maximal and minimal path delays over all path combinations are defined as $E\{d_{\textrm{max}}(\alpha)\}=E\{d_{n}(\alpha)\}$ and  $E\{d_{\textrm{min}}(\alpha)\}=E\{d_{1}(\alpha)\}$, respectively. Finally, using Eq.~\eqref{taumean}, the expected value of differential delay can be derived as, i.e.,
{
\begin{equation}\label{MaxMin}
\bar\tau = E\{\tau(\alpha)\} = \frac{1}{a_{\mathcal M}}\sum_{i=0}^{N-n}(d_{(N-i)}-d_{i+1})C_{N-i-1,n-1}
\end{equation}

\par To derive  further relations, let us define the likelihood that a specific subset $\mathcal M_n(\alpha)$  is chosen by indicate its dependence on the number $n$ of parallel paths. Since only one out of $a_{\mathcal M}$ path combinations is randomly selected for transmission, all path sets and all paths of a set occur with the same probability.
Let us denote the probability of an arbitrary paths combination as $P(\mathcal M_{n}(\alpha))=\frac{1}{a_{\mathcal M}}  \equiv P'(\alpha, n) $, whereby

\begin{equation}\label{combProb}
P'(\alpha, n)=\prod_{m=1} ^{n} P_{l_m}(\alpha) = (P_{\mathcal M })^n   = \frac{1}{C_{N,n}},
\end{equation}
where $P_{\mathcal M }$ is the probability that an arbitrary path $\mathcal P_l \in G_{\mathcal P}$ is selected for transmission and collected in $\mathcal M_n(\alpha)$.  
}


{\par Generally, each element $d_l$ from $\vv d$ can have value equal to values of their direct neighbors, i.e., $d_{l-1}$ and $d_{l+1}$. 
Thus, vector $\vv d$ can contain $D_{\textrm{min}}$ and $D_{\textrm{max}}$  elements with minimal and maximal path delays, i.e., $d_1=d_2=...=d_{D_{\textrm{min}}}$ and $d_{N}=d_{N-1}=...=d_{N-D_{\textrm{max}}+1}$, respectively. Therefore, there maybe many more path combinations $\mathcal M_{n}(\alpha)$ yielding a maximal differential delay defined with Eq.~\eqref{UpperDelay}
\begin{equation}\label{UpperDelay}
\tau_{\textrm{up}}=d_{N}-d_1 \equiv d_{N-D_{\textrm{max}}+1}-d_{D_{\textrm{min}}}
\end{equation}
The occurrence probability of  $\tau_{\textrm{up}}$ is 
\begin{equation}\label{ProbLongest3}
P_{\textrm{up}}=\!\!\!\!\sum_{j=max\{N-D_{\textrm{max}}+1,\atop n\}}^{N}\!\!\!\! \!\!\sum_{i=1}^{min\{D_{\textrm{min}},\atop j-n+1\}}\!\!\!\!\!\!P_i \cdot P_{j}\cdot\!\!\!\!\!\!\!\! \sum_{\alpha=1}^{C_{j-i-1,n-2}}\!\!\!\!P'(\alpha, n-2),
\end{equation}
}{
where for a path combination $\mathcal M_{n}(\alpha)$ we assume, that the $k^{th}$ path  $\mathcal P_{l_{k}}$   with maximal delay is mapped to one of the $D_\text{max}$ paths  $\mathcal P_j \in G_{\mathcal P}$, i.e, $d_{l_{k}}(\alpha) = d_j $ for $j = N-D_\text{max}+1,...,N$. But if  $n >  N-D_\text{max}+1 $,  the mapping of the path with maximal delay has to be limited to $j \ge  max\{n, N-D_\text{max}+1 \}$. Furthermore, its path with minimal delay, $d_{l_1}(\alpha)$, is mapped to one out of $D_\text{min}$ paths  $\mathcal P_i \in G_{\mathcal P}$ for $i = 1,2,...,D_\text{min}$. Otherwise for a selected longest path $\mathcal P_j $, the first  path of a set of $n$ path  is restricted to paths $\mathcal P_i \in G_{\mathcal P}$ with index $i \le j-n+1$, thus the mapping  of  $d_{l_1}(\alpha) = d_i $ is restricted to the range $i \le min \{D_\text{min}, j-n+1 \}$. Finally, there are $C_{j-i-1,n-2}$ possible path combinations, whose probability $P'(\alpha, n-2) =  (P_{\mathcal M })^{n-2}$  follows from Eq.~\eqref{combProb} and is independent of $\alpha$. Furthermore, the shortest and longest paths  in each combination $\mathcal M_{n}(\alpha)$,  i.e., one or more paths with delays $d_1$  and $d_{N}$ from $\vv d$, respectively, occur with the same probability $P_{i} =  P_{j}= P_{\mathcal M }$  in all combinations. Thus  Eq.~\eqref{ProbLongest3} can be simplified as follows
}  {
\begin{equation}\label{ProbLongest4}
P_{\textrm{up}}=\frac{1}{a_{\mathcal M}}\sum_{j=max\{N-D_{\textrm{max}}+1,\atop n\}}^{N}\sum_{i=1}^{min\{D_{\textrm{min}},\atop j-n+1\}} C_{j-i-1,n-2}
\end{equation}
  
\par When delay vector $\vv d$ contains only two elements with maximal and minimal delay, i.e., $D_{\textrm{min}}=1$ and $D_{\textrm{max}}=1$, 
}
{
the summation is omitted in  Eq. \eqref{ProbLongest4}, which yields
\begin{equation}\label{ProbLongest2}
P_{\textrm{up}}=\frac{C_{N-2,n-2}}{a_{\mathcal M}} =\frac{C_{N-2,n-2}}{C_{N,n}} = \frac{n(n-1)}{N(N-1)}
\end{equation}
} {
Similar to derivation of Eq. \eqref{ProbLongest4}, we can additionally derive an occurrence probability of path combinations, where predefined paths $\mathcal P_{y}$ and $\mathcal P_{x}$ are assumed as longest and shortest paths, respectively, resulting in differential delay $\tau_{x,y}=d_y-d_x$. That is equivalent to the special case, where $D_{\textrm{min}}$ and $D_{\textrm{max}}$ are irrelevant for probability calculation. Thus, the occurrence probability of the combinations, which contain certain paths $\mathcal P_{y}$ and $\mathcal P_{x}$ as the longest and the shortest paths, respectively, is 
     \begin{equation}\label{ProbDelay}
       P_{{x,y}}=\frac{C_{y-x-1,n-2}}{C_{N,n}}
     \end{equation}
However, when, in a network with $N$ parallel paths, the paths $\mathcal P_{y}$ and $\mathcal P_{x}$ have delays $d_y<d_{N}$ and $d_x>d_1$, $y<N-D_{\textrm{max}}+1$, $x>D_{\textrm{min}}$, $y-x-1\geq n-2$, the combinations with $\mathcal P_{y}$ and $\mathcal P_{x}$ as the longest and shortest paths have  occurrence probability smaller or equal to the occurrence probability of path combination with maximal differential delay, i.e., $P_{\textrm{up}} \geq P_{{x,y}}$. On the other hand, there can be other path combinations, which do not contain paths $\mathcal P_{y}$ and $\mathcal P_{x}$  as the longest and shortest paths, but result in the same differential  delay value $\tau_{x,y}$. Thus, based on Eq. \eqref{ProbLongest4}, we can claim that the maximal differential delay $\tau_{\textrm{up}}$ has the larger occurrence probability than any other lower value of differential delay $\tau_{x,y}$ only if $P_{\textrm{up}}>0.5$, which is valid for a large values of $D_{\textrm{min}}$ and $D_{\textrm{max}}$.
}

\subsection{Impact of coding redundancy on path analysis}

\par As previously mentioned, $k$ parallel lanes can be coded into $n=k+r$ data blocks, and thereafter transmitted over $n$ paths. Thus, network routing redundancy can be directly related to the coding redundancy. Due to the fact, however, that the destination needs to buffer only $k$ out of $n=k+r$ data blocks from the same generation for the decoding start, while any $r$ data blocks arriving later can be ignored, the application of the previous analysis is not straightforward in this case, and needs a few modified expressions of the expected values of the differential delay.

\par Let us assume that all $n=k+r$ parallel paths are selected randomly with the same probability. {For each path set $\mathcal M_{n}(\alpha)$ the corresponding delays are sorted as $ d_{l_1}(\alpha)\le d_{l_2}(\alpha)\le...\le d_{l_{k}}(\alpha)\le d_{l_{k+1}}(\alpha)\le ...\le d_{l_m}(\alpha)\le...\le d_{l_{k+r}}(\alpha)$, $d_{l_m}(\alpha)=\sum_{i}  d_i \delta_{\alpha}(i,m)$, $d_i$ from $\vv d$, $1\leq m\leq n$.} For successful decoding, the receiver needs to buffer the data blocks arriving from any $k$  paths with least delays, i.e., $d_{l_m}(\alpha), m=1,2,...,k$. Thus, in a parallel network system with RLNC, the sender can in fact arbitrarily assign any of the $n$ paths, irrespectively of their related path delays, which eliminates need for routing and complex cross-layer design.

{In a parallel network system with linear network coding and $r$ redundant paths, where the path set $\mathcal M_n(\alpha)$ of $n=k+r$ paths is randomly chosen, a maximal path delay which has to be taken into account at the decoder is bounded as follows
\begin{equation}\label{newMaxBound}
d_{k}\le d_{\textrm{max}}(\alpha,r) \le d_{N-r}
\end{equation}

The detailed derivation of Eq. \eqref{newMaxBound} is in Appendix.}
With Eqs.~\eqref{newMaxBound},~\eqref{ProbTi} and~\eqref{ExpectedTi}, {an expected maximal path delay} follows to be
\begin{equation}\label{newMax}
E\{d_{\textrm{max}}(\alpha)\}(r)\!\!=\!\!\!\sum_{l=k}^{N-r}\!\!p_l(k)\cdot d_l\!\!=\!\!E\{d_{k}(\alpha)\}\!\!=\!\!E\{d_{n-r}(\alpha)\}
\end{equation}

\par {
Similar the expected minimal path delay can be calculated as $E\{d_{\textrm{min}}(\alpha)\}(r)=E\{d_{1}(\alpha)\}$. Finally, with Eq.~\eqref{newMax} and using the fact that the number of paths is now given by $n=k+r$ the expected value of differential delay follows to be
\begin{equation}\label{newMaxMin}
\bar\tau(r)=E\{d_{n-r}(\alpha)\}-E\{d_{1}(\alpha)\},
\end{equation}
where $\bar\tau(0)=\bar\tau$ simplifies to the result given by Eq.~\eqref{MaxMin}.
}

As a result, the maximal differential delay is
\begin{equation}\label{UpperDelayR}
\tau_\text{up}(r)=d_{N-r}-d_1
\end{equation}

{Due to the fact, that the delay vector $\vv d$ can contain $D_{\textrm{max}}$ and  $D_{\textrm{min}}$ elements $d_l$, which are equal to $d_{N}$ and $d_1$, respectively, it is advantageously to consider a special case covered by   Eq.(\ref{UpperDelayR}). If $r < D_{\textrm{max}}$ the maximal relevant delay is given by $d_{N-D_{\textrm{max}}+1} = d_{N-r}  =  d_N$, and $\tau_\text{up}(r)= d_{N-D_{\textrm{max}}+1} - d_1$ is  independent of $r$. Thus, an equivalent form of  Eq.(\ref{UpperDelayR}) is given by $\tau_\text{up}(r)=d_{N-max\{ r, D_{\textrm{max}}-1\}}-d_1 $. Using this result and similar to Eq.~\eqref{ProbLongest3}, the occurrence probability of maximal differential delays $\tau_{\textrm{up}}(r)$ is given by Eq.~\eqref{ProbLongestR3}.
\begin{equation}\label{ProbLongestR3}
\begin{split}
P_\text{up}(r)&=\!\!\sum_{j=max\{N-max(r,D_{\textrm{max}}-1),\atop n-r\}}^{N-r} P_{j}\!\!\sum_{i=1}^{min\{D_{\textrm{min}},\atop j-n+r+1\}} P_i  \cdot \\
&\cdot\!\!\!\!\sum_{\beta=1}^{C_{N-j,r}} \!\! P'(\beta, r)\!\!\!\! \sum_{\alpha'=1}^{C_{j-i-1,n-r-2}}\!\!P'(\alpha', n-r-2),
\end{split}
\end{equation}

Due to the $r$ redundant paths, only $k=n-r$ paths are effectively used and the maximal used path delay is determined by $d_{N-max\{ r, D_{\textrm{max}}-1\}}$, as discussed above. Thus in contrast to Eq.~\eqref{ProbLongest3} the $k^{th}$ path  $\mathcal P_{l_{k}}$ with maximal delay is mapped to one path  $\mathcal P_j \in G_{\mathcal P}$ within the range given by   $d_{l_{k}}(\alpha) = d_j $ for $j =  max\{n-r,  N-max\{ r, D_{\textrm{max}}-1\} \}   ,...,N-r$. 
Similar, the path with minimal delay, $d_{l_1}(\alpha) $,  is mapped to one of the paths  $\mathcal P_i \in G_{\mathcal P}$ out of   the range $i = 1,..., min \{D_\text{min}, j-n+r+1 \}$. In accordance with our assumptions, the $r$ redundant paths have a delay larger or equal than the $k^{th}$ path. Now depending on the mapping of the $k^{th}$  path to one path  $\mathcal P_j $, the redundant paths  $\mathcal P_{l_{k+1}}(\alpha),...,\mathcal P_{l_{k+r}}(\alpha)$  will be mapped to paths  $\mathcal P_l  \in G_{\mathcal P}$ with delays $  d_l $ out of the range $l = j+1,...,N$. Thus there are $C_{N-j,r}$ path combinations to achieve these mapping, which are combined in subset $\mathcal M_{n}(\beta)$. Due to  Eq.  \eqref{combProb}, the  probability for the set of $r$ redundant path can be written as $ P'(\beta, r)= ( P_{\mathcal M })^r $.  Similar to Eq.~\eqref{ProbLongest3}, for the remaining $n-r-2$ paths there are $C_{j-i-1,n-r-2}$  combinations whose paths are combined in subset $\mathcal M_{n}(\alpha')$, where the probability for a set of $n-r-2=k-2$  paths  can be written as $ P'(\alpha', n-r-2)= ( P_{\mathcal M })^{k-2} $.  Thus the probability  $P_{\textrm{up}}(r)$ simplifies to   

\begin{equation}\label{ProbLongestR4}
\begin{split}
P_\text{up}(r)=\! 1 / a_{\mathcal M}  \! \! \! \! \!\! \! \! \! \! \! \! \! \! \! \! \!\! \! \! \!     \sum_{j=max\{N-max(r,D_{\textrm{max}}-1),\atop n-r\}}^{N-r}\!\!\!\!\!\! \! \! \! \! \! \! \! \! \! \! \!\! \! \! \!   C_{N-j,r}   \cdot  \! \! \! \! \!    \sum_{i=1}^{min\{D_{\textrm{min}},\atop j-n+r+1\}}\!\!\!\!\!\!\! C_{j-i-1,n-r-2},
\end{split}
\end{equation}


For the case, there is only one path with maximal delay, $d_{N-r}$ and only one path with minimal delay, $d_1$, i.e $D_{\textrm{max}}=D_{\textrm{min}}=1$, the probability of occurrence of a maximal differential delay can be derived from Eq.~\eqref{ProbLongestR4} as
\begin{equation}\label{ProbLongestR2}
P_\text{up}(r)= \frac{C_{N-r-2,n-r-2}}{C_{N,n}}=\prod_{i=0}^{r+1}\frac{(n-i)}{(N-i)}
\end{equation}
}
\par Compared to the system without redundancy, the receiver experiences \emph{a reduction of a maximum differential delay} in the network $\boldsymbol\Delta^\text{max}$, expressed as follows
\begin{equation}\label{reduction}
\boldsymbol\Delta^\text{max}=\frac{d_{N}-d_{N-r}}{d_{N}}
\end{equation}
Additionally, the expected value of differential delay can be reduced by the ratio $\boldsymbol\Delta^{\tau}$ determined by Eq.~\eqref{reductionDD}.
\begin{equation}\label{reductionDD}
\boldsymbol\Delta^{\tau}=\frac{\bar\tau-\bar\tau(r)}{\bar\tau}
\end{equation}
\par On the other hand, sending of $r$ redundant data flows over additional $r$ paths presents a capacity and transmission overhead, which we define and later numerically evaluate as
 \begin{equation}\label{overhead}
\boldsymbol {\Theta}=\frac{r}{k}
\end{equation}

\begin{figure*}[!t]
\subfigure[]{
\includegraphics[width=0.86\columnwidth]{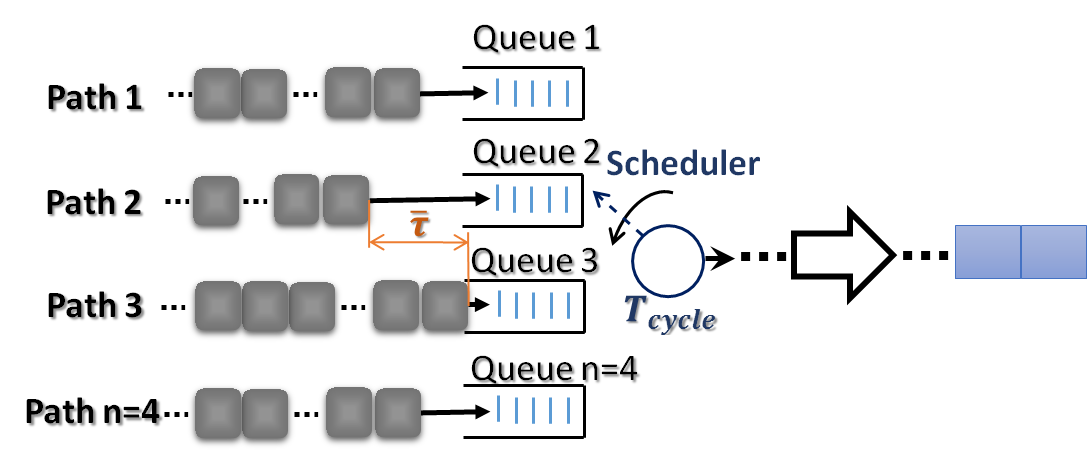}
\label{deskew}
}\hspace{5mm}
\subfigure[]{
\includegraphics[width=0.99\columnwidth]{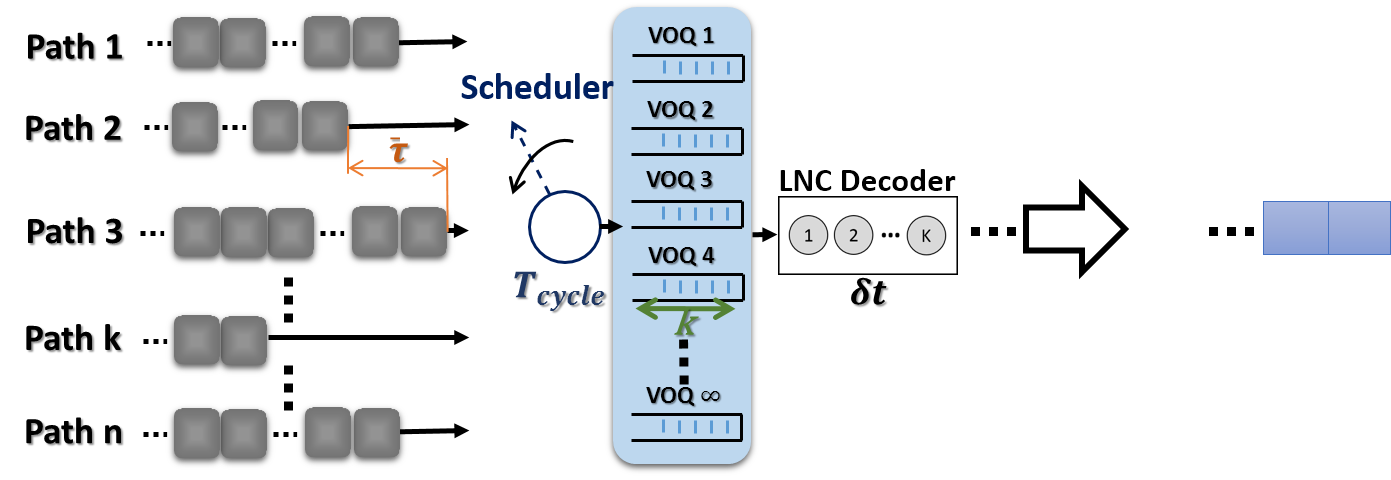}
\label{decodbuffer}
}
\vspace{-0.2cm}
\caption{Buffer models at the receiver. (a)  Deskew buffer per lane; (b) Decoding buffer. }\label{buffers}
\vspace{-0.3cm}
\end{figure*}

\subsubsection{Discussion on path failures and packet loss}
{In contrast to coded parallel transmission without redundancies, the utilization of redundant paths does not only provide a reduction of differential delay, but also can increase robustness of system regarding packet, i.e., coded data blocks loss and path failures, when the number of lost packets per generation $f_\text{pack}$ and number of failed paths $f_\text{path}$ is less or equal to the number of coding and path redundancies $r$, respectively. However, we need to consider a possible impact of any fault on resulting differential delay. 

With path failures, while  $f_\text{path}$ available paths in the network fail, the source can utilize only $N-f_\text{path}$ of the remaining paths. Thus, the expected differential delay can be calculated by Eq. \eqref{newMaxMin} under consideration of reduced vector $\vv d$, which contains now $N-f_\text{path}$ elements. 

In case of packet loss, while $f_\text{pack}\leq r$, we consider a worst case, whereby packet loss can occur on any utilized path. In that worst case scenario, the maximal possible differential delay needs to be considered for system design to prevent possible retransmission. Thus, the mean differential delay in system with packet loss can be calculated with  Eq. \eqref{ExpectedTi} as
\begin{equation}\label{newMaxMinFail}
\bar\tau(r)=E\{d_{k+r}(\alpha)\}-E\{d_{1}(\alpha)\},
\end{equation}
}

\subsection{Analysis of receiver queue size}

\par In a parallel network system without RLNC, we refer to the required queue size, as the \emph{deskew buffer} (Fig.~\ref{deskew}). In the system with RLNC, on the other hand, the buffer architecture is based on the virtual output queue (VOQ), referred to as the \emph{decoding} buffer (Fig.~\ref{decodbuffer}). 
 Let us denote the delay difference between the longest and an arbitrary path in a path set $\mathcal M_{k}(\alpha)$ as $\tau_m(\alpha)\!\! =\!\!d_{l_{k}}(\alpha)-d_{l_m}(\alpha)$, $m\!\! =\!\! 1,2,...,k$, 
{
where $\tau(\alpha)\!\! =\!\! \tau_1(\alpha) $ refers to the largest differential delay within the path set and can be determined with Eq.~\eqref{tau}.

\par To analyze the effect of differential delay for $n$ input lanes, in this section we first assume a fixed optimal path pattern $\mathcal M_{n}(\alpha_{\textrm{opt}})$ with delays $d_{l_m}(\alpha_{\textrm{opt}})$ ordered as $ d_{l_1}(\alpha_{\textrm{opt}}) \le d_{l_2}(\alpha_{\textrm{opt}}) \le...\le d_{l_{n}}(\alpha_{\textrm{opt}}) $.  
}  Here, the largest differential delay within the optimal path set is denoted as $\tau(\alpha_{\textrm{opt}})$. Generally, the expected value of the buffer size strongly depends on the paths chosen from vector $\vv d$.

However, due to the definition of $\tau_m(\alpha)$, the accumulated differential delay between all selected paths is given by $\sum^{k}_{m=1} \tau_m(\alpha)$. In contrast to randomly selected paths, the optimal path set will minimize the differential delay between selected paths and, thus, the buffer size. As a result, the optimal path combination is given by $\mathcal M_{n}(\alpha_{\textrm{opt}})=\mathcal M_n(\alpha)$, when $min_{\alpha}\{  \sum^{k}_{m=1} \tau_m(\alpha) \}$.


\par We assume an idealized scheduler, whereby, in steady state, we assume a deterministically distributed arrival process. 
{The assumption that arrivals are deterministic is because $n$ paths are selected in the optical layer to transport a single Ethernet frame. For instance, for a 1500 byte frame and $n=4$ parallel lanes, 47 blocks will arrive in succession on each of the parallel lanes at the receiver, which makes it deterministic. }

For $n$ input lanes, the idealized scheduler runs $n$ time faster, such that during a full cycle time $\Delta t_{cycle}$ a total of $n$ data blocks are forwarded and processed every \emph{tu} as defined next 
\begin{equation}\label{cycle}
 \Delta t_{cycle}(n) = n\cdot (t_{poll}+t_f) \leq 1 [tu],
\end{equation}
where  $t_f$ is the mean forwarding time of a data block and $t_{poll} \ll t_f $ is the polling and processing time. 

\subsubsection{Deskew Buffer Size}

\par
This architecture reflects the multi-lane Ethernet technology, where the number of parallel paths in $\mathcal M_{n}(\alpha_{\textrm{opt}})$ is equal to the original number of virtual lanes used in the sender, i.e. $n=k$. In terms of buffer sizing, the worst case scenario occurs when the receiver needs to deskew the differential delay
{between the lanes with largest and smallest delay $\tau(\alpha_{\textrm{opt}}) = \tau $ given by Eq.~\eqref{tau}, where the path set index $\alpha_{\textrm{opt}}$ is neglected to simplify the notation. This delay,
}
 if measured in multiples of \emph{time units}, requires a buffer size of $(\tau)$ for the shortest lane. The reordering and re-serialization process is undertaken by the scheduler, which requires an additional buffer place $1 [tu]$ per lane to enable the processing after the cycle delay $\Delta t_{cycle}(k)$. The latter is a simple model for Ethernet's multi-lane deskewing mechanism using data block markers. Thus, the queue size required for the lane related to the shortest path is $(\tau + 1)$. To allow for any arbitrary pattern of differential delay, the input buffer size is the same for each lane, which corresponds to a classical design principle used in parallel hardware architectures. Consequently, the total buffer size can be expressed as
\begin{equation}\label{Multipath}
\Omega_\text{ML} = k  \cdot \left(\tau +  1 \right)
\end{equation}

 \subsubsection{Decoding queue size}
\par The first data blocks from any $n$ lanes and originated from first generation created at the RLNC sender typically arrive at the receiver at different times. To analyze the effect of differential delay for $n=k$ input lanes, we first assume a fixed path pattern $\mathcal M_{n}(\alpha)$ with delays $d_{l_m}(\alpha)$ ordered as $ d_{l_1}(\alpha)\le d_{l_2}(\alpha)\le...\le d_{l_{k}}(\alpha)$. 

The scheduler has to poll $k$ input lanes and as soon as the first data block of a new generation arrives and that on a lane corresponding to the shortest path, the scheduler forwards the data block to a newly created virtual output queue (VOQ). In the initial phase, the decoding commences when $k$ data blocks of the first generation arrive. Consequently, the data blocks from the same generation arriving from the $k-1$ shorter paths have to be buffered in the decoding buffer until the $k^{th}$ data block from the same generation is received. Thus, before the last $k^{th}$ packet from the first generation arrives, $ \Omega_\text{ini}(\tau)$ packets from other $k-1$ lanes must be buffered. The queue size during the initial phase is thus
{
\begin{equation}\label{initialBuffer}
   \Omega_\text{ini}(\tau) = \sum^{k}_{m=1} \tau_m(\alpha)
\end{equation}
where the explicit dependence on the path set index $\alpha$ is omitted.   }    {Note, as soon as there is no differential delay, i.e., $\tau=0$ $tu$, the initial size of the decoding queue Eq.\eqref{initialBuffer} becomes zero.} This is due to the fact that data blocks are immediately transferred by the scheduler during the subsequent cycle.

\par For the deterministic arrival process, steady state is reached after the first generation completes. Here, the amount of data blocks forwarded in every decoding interval $t_{\delta}$ is
\begin{equation}\label{interBuffer}
\Omega_{\delta}(t_{\delta}) =    \left\lfloor\frac{t_{\delta}}{t_\text{poll}+t_\text{f}}\right\rfloor
\end{equation}


\par In steady state, the decoding process finishes after the decoding interval $t_{\delta}$, and all data blocks from one complete generation leave the decoding buffer. To avoid idle periods, a new decoding process should immediately commence after previous decoding cycle is finished. In the best case, the scheduler transfers a new generation to the decoder every $\Delta t_\text{cycle}(k)$, i.e.   $t_{\delta}\le \Delta t_\text{cycle}(k)$. On the other hand, one data block forwarded to the decoding buffer can complete a generation in case of $\tau>0$. Thus, one data block is sufficient to trigger the subsequent decoding process. For this purpose, the decoding interval $t_{\delta}$ can take a value in the range $\frac{\Delta t_\text{cycle}(k)}{k} \le t_{\delta} \le \Delta t_\text{cycle}(k) $. Although a short decoding interval $t_{\delta}$ will release complete generations very rapidly, a new decoding interval can only start after a new generation has reached the decoding buffer, which in worst case can be as long as $\Delta t_\text{cycle}$. Thus, the longest possible decoding interval is the best estimate, and defined as $t_{\delta}= \Delta t_\text{cycle}(k)$. As a result, the decoding queue size can be calculated by combining  Eq.\eqref{cycle}, Eq.\eqref{initialBuffer} and Eq.\eqref{interBuffer} as
{
\begin{equation}\label{decoding}
\Omega_\text{LNC} =  \Omega_{\delta}({\Delta t_\text{cycle}(k)}) \!+ \! \Omega_\text{ini}(\tau) = k + \!\! \sum^{k}_{m=1} \!\!  \tau_m(\alpha)
\end{equation}
}

We next analyze the VOQ buffer architecture with redundancy, i.e., $n=k+r$, $r\geq0$, and random path selection. This is to show that under regular operation mode, i.e., in absence of path failures, the redundant paths not only can reduce the differential delay $\bar\tau(r)$, as discussed earlier, but also the buffer size required. 
Similarly to the previous analysis, derived from Eq.~\eqref{decoding} with Eq. \eqref{newMaxMin},  the average size of the decoding buffer is
\begin{equation}\label{decodingExpexact}
\Omega^{^\text{R}}_\text{LNC}(r)\!=\!k\!+\!E\{\! \sum^{n-r}_{m=1}\!\!\!\! \tau_m(\alpha,r) \} \!=\! k\!+\!\!\! \sum^{n-r}_{m=1}\!E\{ \tau_m(\alpha,r) \},
\end{equation}
where $E\{ \tau_m(\alpha,r)\}$ is an expected value of the so-called \emph{path-specific differential delay}, defined below.  {To derive this value, for each path set $\mathcal M_{n}(\alpha)$, 
we define the difference between  neighboring paths $\mathcal P_{l_{\nu+1}}(\alpha)$ and $\mathcal P_{l_\nu}(\alpha)$, as $\Delta \tau_\nu(\alpha,r) =d_{l_{\nu+1}}(\alpha)-d_{l_\nu}(\alpha)$, $\nu = 1,2,...,n-r-1$.
{ 
Following Eq.~\eqref{ExpectedTi} and using $\tau_m(\alpha,r)=d_{l_{n-r}}(\alpha) -d_{l_{m}}(\alpha)= \!\!\sum_{\nu=m}^{n-r-1} \!\! \Delta \tau_\nu(\alpha,r)$     }
the expected value of path-specific differential delay is given by
\begin{equation}\label{ExpectedTaui}
E\{\tau_m(\alpha,r)\} \!\!  = \!\!  \sum_{\nu=m}^{n-r-1} \!\! E\{ \Delta \tau_\nu(\alpha,r) \}, 1\leq m \leq n-r-1
\end{equation}
and $E\{ \tau_{n-r}(\alpha,r) \}=0$, where 
{  $E\{\Delta \tau_\nu(\alpha,r)\}$ }     is the expected value of the delay difference between two sequent lanes from $\mathcal M_{n}(\alpha)$. To get more insights into the general behavior of the path selection, these values are assumed to be identical,
{  i.e. $E\{\Delta \tau_\nu(\alpha,r)\}\!\! =\!\! E\{\Delta \tau(\alpha,r)\} \!\!=\!\! \Delta \bar \tau$. Under this assumption, the expected value of the largest differential delay   }
follows to be $E\{\tau_1(\alpha,r)\}\!\!=\!\!\bar\tau(r)\!\!=\!\!(n-r-1)\Delta \bar \tau$, which allows to derive $ \Delta \bar \tau\!\!=\!\!\frac{\bar\tau(r)}{n-r-1}\!\!=\!\!\frac{\bar\tau(r)}{k-1}$. Furthermore, with $k\!\!=\!\!n\!\!-\!\!r$ Eq.~\eqref{ExpectedTaui} can be simplified to $E\{\tau_m(\alpha,r)\}\!\! =\!\! \sum_{\nu=m}^{k-1} \Delta \bar \tau\!\! =\!\! (k-m) \Delta \bar \tau $. 
Thus, an approximation for Eq.~\eqref{decodingExpexact} follows as
\begin{equation}\label{decodingExp2}
\Omega^{^\text{R}}_\text{LNC}(r) \approx k+\sum^{k}_{m=1} (k-m) \Delta\bar\tau    = k+\bar\tau(r)\cdot \frac{k}{2}
\end{equation}
}
\subsubsection {Discussion on random routing}
The analysis of the routing based on random path selection can be viewed as evaluation of all possible routing schemes in a network. { When no optimization is applied to routing and data blocks are sent over any $n=k$ out of $N$ paths, this results in variations of differential delay. The expected value of the queue size can be derived in closed form. From Eq.~\eqref{MaxMin} and Eq.~\eqref{Multipath} the average size of the deskew buffer is
\begin{equation}\label{ExpMultipath}
\Omega_\text{ML}^\text{R} = E\{ k  \cdot \left( \tau(\alpha) +  1 \right)\} = k  \cdot \left( \bar\tau + 1 \right)
\end{equation}
}
 This allows us to compare both buffer architectures in a generalized statistical way. Thus let us estimate the advantage of the decoding over the deskew based buffer architecture using Eq.\eqref{ExpMultipath} and \eqref{decodingExp2}, if $r=0$ and $\bar\tau(0)=\bar\tau$, which yields
\begin{equation}\label{RNDbuffercompare}
\frac{\Omega^{^\text{R}}_\text{ML}-\Omega^{^\text{R}}_\text{LNC}(0)}{\Omega^{^\text{R}}_\text{ML}} \approx \frac{\bar\tau}{2(\bar\tau+1)} \xrightarrow[\bar\tau \gg 1]{}  \frac{1}{2}
\end{equation}
In conclusion, the decoding, i.e., VOQ-based, buffer architecture has in the average an advantage over the deskew architecture independently of the number of parallel paths used in the network, with an up to $50\%$ of improvement in buffer size in case of large differential delays.

\subsubsection{Analysis of upper and lower bounds of decoding buffer}

\par The upper bound of the decoding buffer, $\Omega^\text{LNC}_\text{up}(r)$, $r\!\!\geq\!\! 0$, can be derived by  considering all $N$ available paths with the corresponding path delays. The worst case for the decoding queue size is defined where $k\!-\!1$ data flows with delays $d_1\leq d_2\leq ...\leq d_{k-1}\}$ from $\vv d$ arrive at the destination and need to be buffered, while the $k^{th}$ data block from each generation required for decoding start always takes a path with the largest path delay $d_{N\!-\!r}$, according to Eq.~\eqref{newMaxBound}. This path set is denoted as $\mathcal M_{n}(\alpha_\text{up})$ and with $\tau_m(\alpha_\text{up})\!\!=\!\!d_{(N\!-\!r\!)_{k}}(\alpha_\text{up})\!\!-\!\!d_{l_m}(\alpha_\text{up})$, where $d_{l_m}(\alpha_\text{up})$ can be $d_{1_1}(\alpha_\text{up})\leq d_{2_2}(\alpha_\text{up})\leq ...\leq d_{(k\!-\!1)_{k\!-\!1}}\leq d_{(N\!-\!r)_{k}}(\alpha_\text{up})$  the upper bound is given as
\begin{equation}\label{LNCUp}
\Omega^{^\text{LNC}}_\text{up}(r)= k + \sum^{n-r}_{m=1} \tau_m(\alpha_\text{up})
\end{equation}
Next, let us establish the absolute lower and upper bounds on Eq.~\eqref{decoding} for arbitrary path sets and topologies.
For the upper bounds, we define that $k-1$ data flows arrive at the destination at the same shortest time and, thus,
~have $\tau_m = \tau_\text{up}, m = 1,2,...,k-1$, with $\tau_\text{up}$ given in Eq.~\eqref{UpperDelay} and buffer upper bound
\begin{equation}\label{LNCup}
\Omega_\text{up}=k+(k-1) \cdot\tau_\text{up}
\end{equation}
For lower bounds, on the other hand, we define that $k-1$ paths have the same maximal delay and one path has a minimal delay, e.g.,$\tau_m = 0, m= 2,...,k$ and  $\tau_1= \tau_\text{up}$. Using Eq.~\eqref{decoding}, the lower bound is given by
\begin{equation}\label{LNClow}
\Omega_\text{low} = k + \tau_\text{up}
\end{equation}

\section{Performance Evaluation}

\par We now present the numerical results and validate the same by simulations. Since we assume a case study of Ethernet-over-optical network, all traffic data blocks can have the same size (66 bits), where last 64b are used for RLNC in a field $F_{2^8}$, with the number of lanes studied is $k=4,8$. The transmission request of each network path requests 10 Gb/s, with the resulting \emph{time unit (tu)} of $6.6 ns$. {A network is modeled as a directed graph with a set of nodes interconnected with 10 parallel links between each pair of nodes. In the physical network, this may imply that each pair of node is connected by a fiber link, and each fiber link carries 10 wavelengths, and thus is able to connect to 10 end-system interfaces at a time, e.g., each at 10Gb/s.} The traffic load is assumed to be 1 data block/\emph{tu} per lane and the network is lossless and fault-free.
\begin{figure}[t]
\includegraphics[width=1\columnwidth]{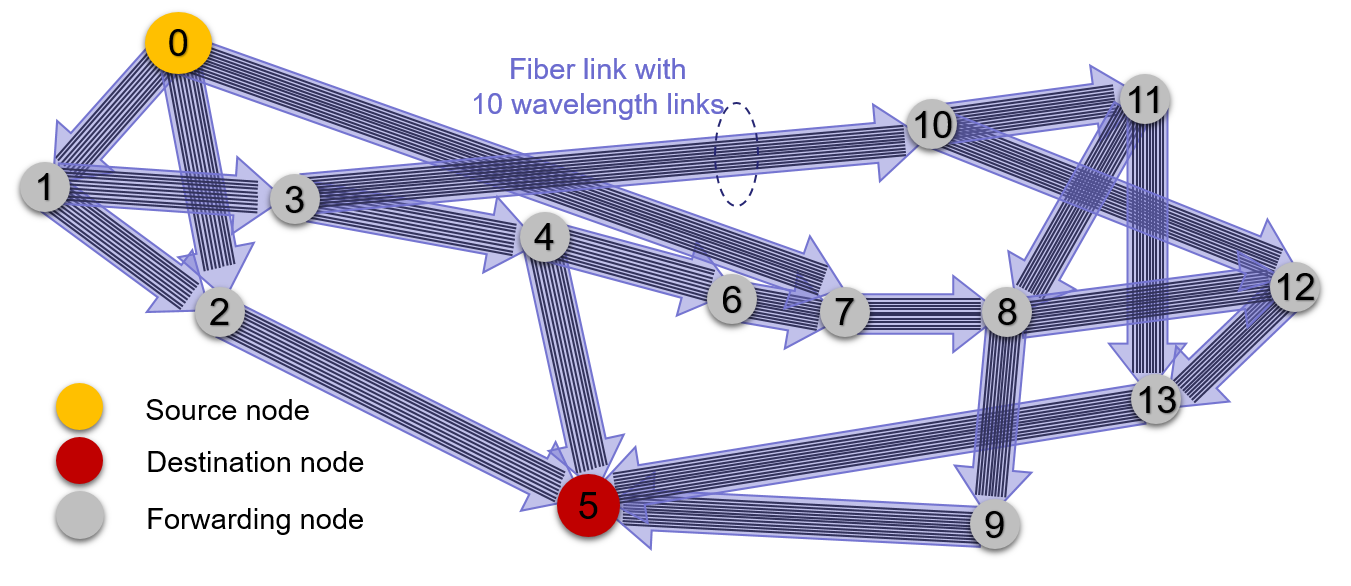}
\caption{nsfNet topology studied.
}\label{topoNSF}
\end{figure}
\begin{table}[!t]
\caption{{Possible optical fiber paths in nsfNet.}}
\label{tab}
\centering
\begin{tabular}{|c|l|}
\hline
$l$ &$\mathcal P_l$\\
\hline\hline
1 & 0-2-5\\
2 & 0-1-2-5\\
3 & 0-7-8-9-5\\
4 & 0-1-3-4-5\\
5 & 0-7-8-12-13-5\\
6 & 0-1-3-10-12-13-5\\
7 & 0-1-3-10-11-13-5\\
8 & 0-1-3-10-11-8-9-5\\
9 & 0-1-3-10-11-8-12-13-5\\
10 & 0-1-3-4-6-7-8-9-5\\
11 & 0-1-3-4-6-7-8-12-13-5\\
\hline
\end{tabular}
\end{table}

\par We analyze two network scenarios. In one scenario, we studied a real world network example, the nsfNet network with 14 nodes and 21 fiber links (Fig.~\ref{topoNSF}). We show the results for the traffic between node $0$ as source and node $5$ as destination, noting that the choice of source and destination points impacts the values of the resulting path delay vector, and is here used for illustration. Each link in the NSFnet exhibits the same delay of $1 tu$. 
{In this topology, there are $11$ different fiber link paths presented in Table \ref{tab} and, since every fiber carries 10 wavelengths, there are $F=110$ possible wavelength paths between source $0$ and destination $5$. Clearly, all wavelength paths within the same fiber link path have the same end-to-end delay. For instance, on path nr. 2 there are 3 fiber hops and all wavelengths using this fiber path have the same end-to-end delay of $3tu$. Between source $0$ and destination $5$, at most $F=25$ wavelength disjoint parallel paths, i.e., $N\leq 25$, can be established simultaneously. This is because if the same wavelengths is chosen on various fiber paths that share the same fiber link there would be a conflict on which of the paths would use that particular wavelength. From the perspective of differential delay, however, only the paths that are wavelength link disjoint are interesting, and we consider only $F'=\frac{F}{10}=11$ wavelength paths in our example. In other words, all the paths listed above, 1,2...11 either allocate a different wavelength, or a fiber link disjoint.  The resulting delay vector is $\vv d'=(2, 3,  4, 4, 5, 6, 6, 7, 8, 8, 9)$. In this scenario, the maximum differential delay $\tau_{\textrm{up}}=d_{F}-d_1$ is $7$ \emph{tu}.} In the second scenario, we assume an abstract network topology that can provide $F'$ fiber paths between source and destination, with the path delays $d'_l$ collected in vector $\vv d'$, each index $l$ corresponding to the path delay, i.e., $d'_l=l$ \emph{tu}. Since our model works with the dynamic delay vector in a network, which depends on topology and network capabilities and is the only factor relevant to the performance analyzed, we validate the analysis in arbitrary networks by using dynamic Monte-Carlo-simulations instead of event simulations, for efficiency. All simulation results are obtained with 95\% confidence interval.
\par The results shown compare three basic methods, i.e.,
\newline 1) LNC-RND. This method corresponds to our model proposed here. Here, all paths are evaluated in a statistical fashion and data blocks can be forwarded over any available paths. In other words, any  $n=k+r$ loop-free paths available between source and destination can be assigned. The decoding buffer size is determined by Eq.~\eqref{decodingExpexact}, while the delay pattern of randomly selected paths is determined by Eqs.~\eqref{ExpectedTi} and ~\eqref{ExpectedTaui}.
\newline 2) LNC-OPT. This is a variation of method 1) with the difference that $n=k$ paths in the network are chosen as "optimal"; here, optimality refers to choosing and allocating those parallel paths that yield the minimum differential delay. This is a common scenario in today's networks, such as equal cost multipath (ECMP) routing. The decoding buffer size is found by Eq.~\eqref{decoding}, while the path delays are known.
\newline 3) ML-OPT. This method does not use RLNC, whereby a set of parallel optimal paths with minimum differential delay is found. The number of paths $n$ chosen equals the number of Ethernet multi lanes $k$, i.e., $n=k$. The deskew buffer is determined by Eq.~\eqref{Multipath}, whereby the differential delay $\tau$ is known and constant.

\begin{figure}[t]
\subfigure[$k=4$]{
\includegraphics[width=1\columnwidth]{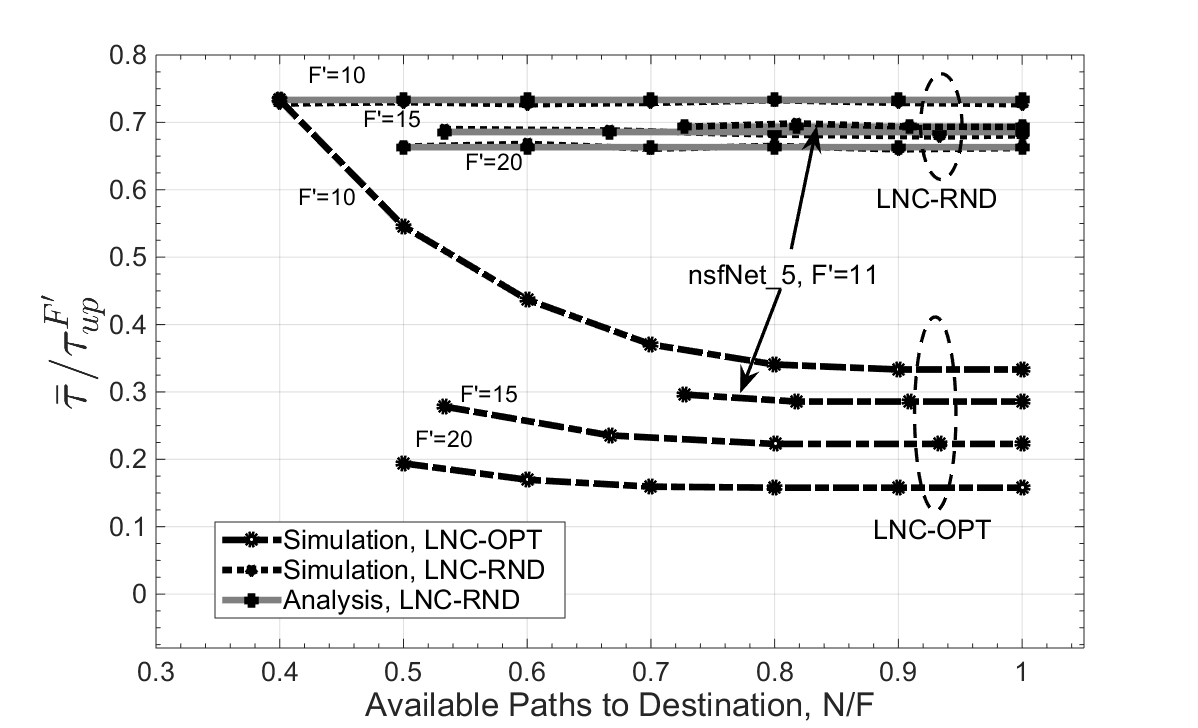}
\label{meanDDk4}\vspace{-0.4cm}
}\\
\subfigure[$k=8$]{
\includegraphics[width=1\columnwidth]{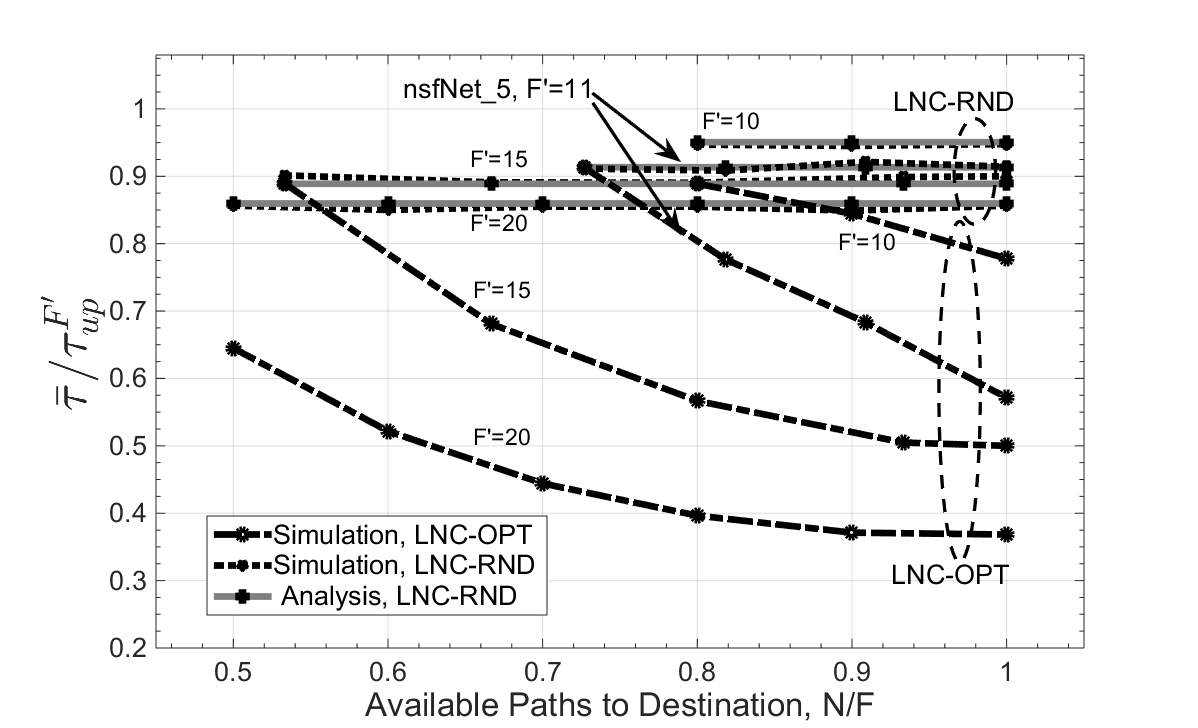}
\label{meanDDk8}
}\vspace{-0.4cm}
\caption{Normalized mean differential delay vs. number of  parallel paths available, with $F'\in\{10,15,20\}$ paths.
}\label{meanDDK48}\vspace{-0.3cm}
\end{figure}

\begin{figure}[t]
\includegraphics[width=1\columnwidth]{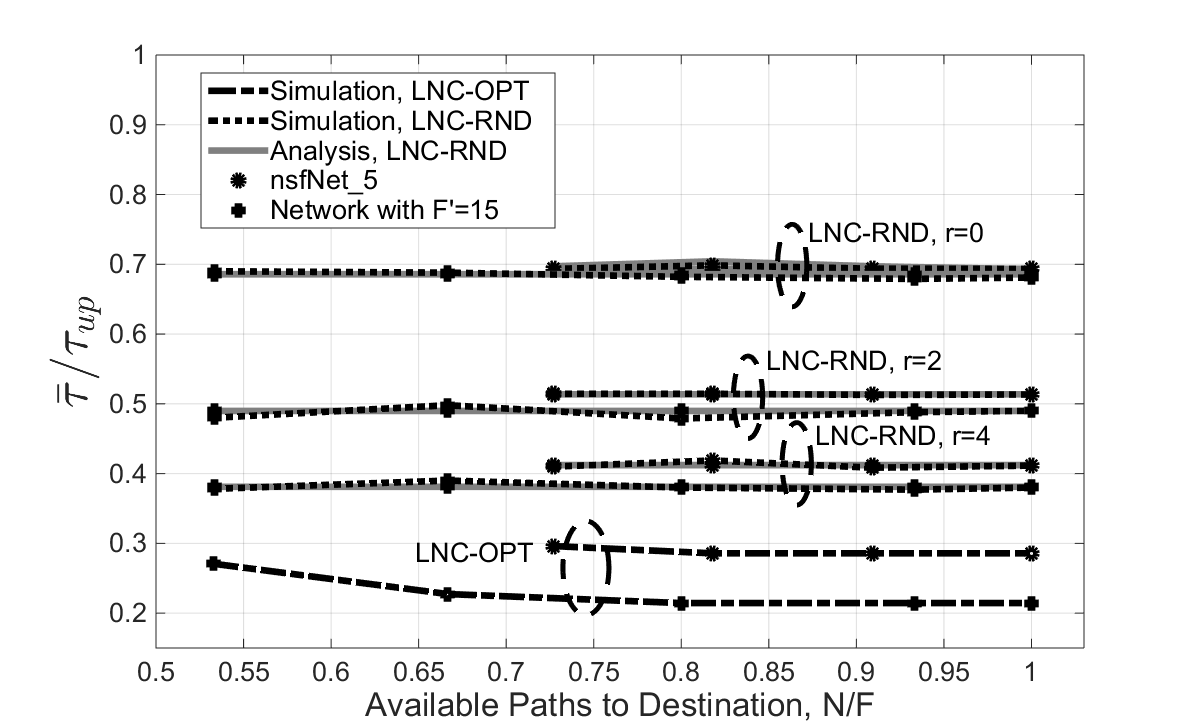}
\caption{Normalized mean differential delay vs. amount of available parallel paths and redundancy in nsfNet$\_5$ and an abstract network with $F'=15$, for $k=4$.
}\label{meanDD413}\vspace{-0.4cm}
\end{figure}

\begin{figure}[!t]
\includegraphics[width=1\columnwidth]{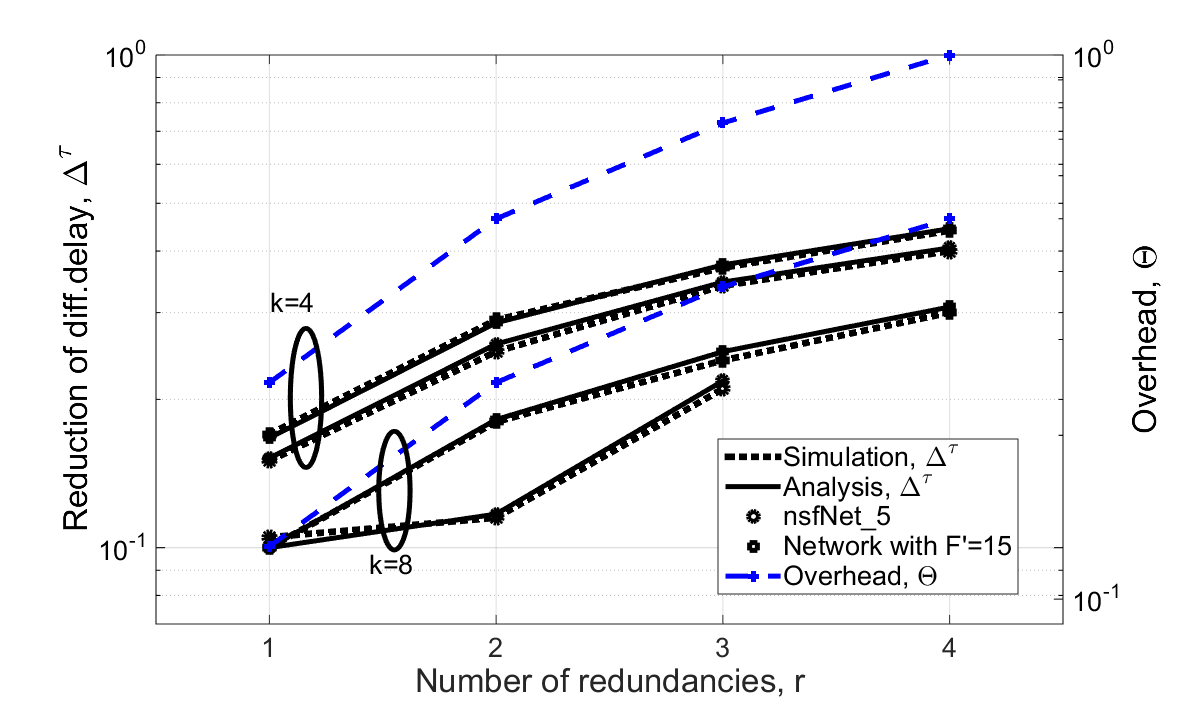}
\caption{Reduction of diff. delay vs. redundancy and generation size for LNC-RND.
}\label{Reduction}
\vspace{-0.5cm}
\end{figure}

\par The first set of results analyses the differential delay as a function of the total number of paths available $N$, out of $F$ possible paths, resulting in $a_\mathcal F = C_{F,N}$ possible path combinations. All results shown for $N/F$ are averaged over all $a_\mathcal F$ combinations and can be expressed as $E(f(n), N)=\frac{1}{a_\mathcal F}\sum_{\alpha=1}^{a_\mathcal F}f_{\alpha}(n)$, where $f(n)$ is differential delay expressed for defined parameters $n$ and $N$, whereby $f_{\alpha}(n)$ relates to a path combination $\alpha$ out of all $a_\mathcal F$ possible  sets of $N$ paths. The mean differential delay is normalized by the maximum differential delay $\tau_{\textrm{up}}$, corresponding to $\tau^{F'}_{\textrm{up}}=(F'-1) tu$ in the abstract network, e.g., $\tau^{15}_{\textrm{up}}=14 tu$ with $F'=15$, and $\tau_{\textrm{up}}=7$ for nsfNet$\_5$. As shown in Fig.~\ref{meanDDK48}, the number of paths available $N$ generally does not affect the expected values of differential delay for LNC-RND. In contrast, the mean value of differential delay decreases with number of available paths $N$ in LNC-OPT  method and can be reduced significantly,  especially for a smaller level of end-system parallelism (number of lanes $k$) and a larger number of all possible paths $F$. Generally, the mean differential delay nears the upper bound with decreasing number of existing paths $F$ and increasing $k$, while optimal routing methods outperform the random path choice method. Although the delay vector $\vv d$ of nsfNet$\_5$ differs from delay vectors of abstract networks, the normalized mean differential delay of nsfNet$\_5$ is between normalized expected values of differential delay of abstract networks with $F'=10$ and $F'=15$ (Fig.~\ref{meanDDK48}).The simulation results match the theoretical results determined by Eqs.~\eqref{tau} and~\eqref{MaxMin}.

\par Fig.~\ref{meanDD413} analyses the same scenarios with redundancy. As expected, the normalized mean differential delay decreases and nears the optimal value with growing number of redundancies and reaches around $39\%$ and $41\%$ of the maximum possible differential delay $\tau_{\textrm{up}}$ in an abstract network with $F'=15$ existing paths and  nsfNet$\_5$, respectively. Here, the theoretical results were calculated with Eq. ~\eqref{newMaxMin}, and match simulations.

\par Fig.~\ref{Reduction} shows the reduction of differential delay and the corresponding transmission overhead (Eq.~\eqref{reductionDD} and Eq.~\eqref{overhead}) as a function of generation size, i.e., number of lanes $k$ and redundancy $r$. All $N=F=110$ paths in nsfNet$\_5$ and $F=150$ paths in the abstract network are available. The simulations and analysis show that an increased number of redundancies $r$ reduces the differential delay, and rapidly increases the transmission overhead, of about 30\%, for $r=3$.

\par
\begin{figure}[t]
\includegraphics[width=1\columnwidth]{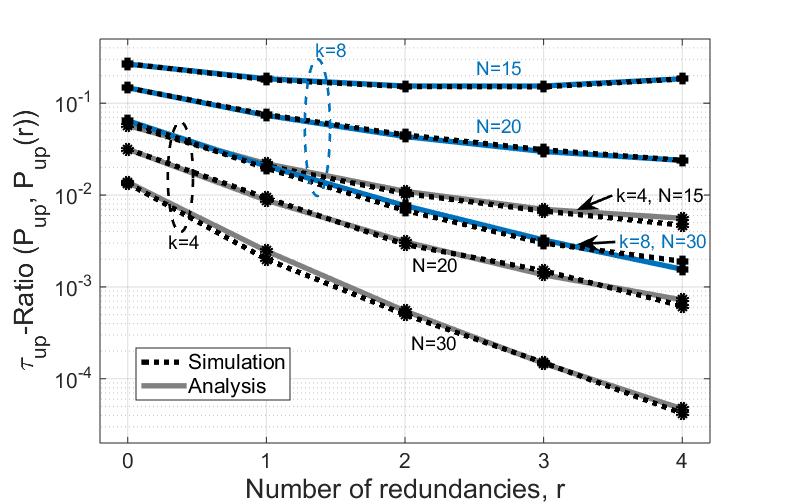}
\caption{The occurrence rate of a maximal differential delay for LNC-RND, when $D_{\textrm{min}}=D_{\textrm{max}}=1$.
}\label{DDupRate48}\vspace{-0.4cm}
\end{figure}

 Fig.~\ref{DDupRate48} shows the occurrence rate of a maximal differential delay $\tau_{\textrm{up}}$ in case of LNC-RND (abstract network), where the path delay vector is defined as discussed above, i.e., $d_l=l$ $tu$, $d_l$ from $\vv d$, $D_{\textrm{min}}=D_{\textrm{max}}=1$, which means there is only one path with a minimal and one path with a maximal path delay per path set. The occurrence rate of $\tau_{\textrm{up}}$ decreases with increasing number of redundancies $r$ and, at the same time, increases with level of end-system parallelism $k$, e.g., $P_{\textrm{up}}\approx0.057$ for $k=4$ and $P_{\textrm{up}}\approx0.27$ for $k=8$ and $r=0$ in a network with $N=15$ parallel available paths. The occurrence probability of a maximal differential delay $P_{\textrm{up}}(0)$ is reduced from around $5.7\%$ to around $1.3\%$, when the number of available paths is doubled, i.e., changed from $N=15$ to $N=30$, respectively, while transmission was established over $n=k=4$ parallel paths. The transmission over $n=k=8$ parallel paths in the same scenario result in a change of $P_{\textrm{up}}(0)$ from around $27\%$ to around $6\%$. This is due to the fact that increase in $F$ results in increase in number of possible paths combinations, which are equally probable. Both simulation and numerical results are based on Eqs.~\eqref{ProbLongest2} and ~\eqref{ProbLongestR2}, while $\tau_{\textrm{up}}(r)$ was defined according to Eq.~\eqref{UpperDelayR}.

\par
\begin{figure}[t]
\includegraphics[width=1\columnwidth]{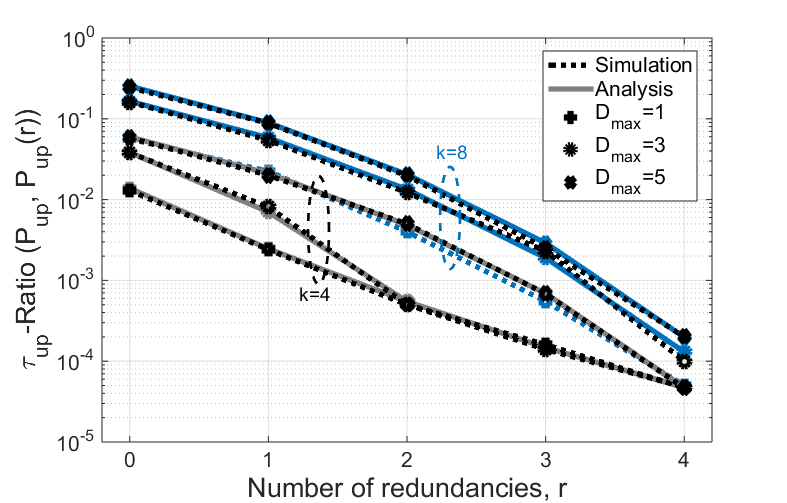}
\caption{The occurrence rate of a max. diff. delay vs. number of paths with max. delay $D_{\textrm{max}}$ in LNC-RND,  $D_{\textrm{min}}=1$.
}\label{DDupRateK3}\vspace{-0.3cm}
\end{figure}

 For a network scenario with $F=N=30$ available paths, and only one path with a minimal path delay, i.e., $D_{\textrm{min}}=1$, Fig.~\ref{DDupRateK3} illustrates the occurrence probability of a maximal differential delay $\tau_{\textrm{up}}(r)$ as a function of number of paths with the same maximal path delay, i.e., $D_{\textrm{max}}$ and the number of redundant paths $r$. In other words, the network studied provides $N-D_{\textrm{max}}$ paths with different delays, $d_l=l tu$, $l=1,..,N-D_{\textrm{max}}$, and $D_{\textrm{max}}$ paths with a maximal delay $d_i=d_{N}=Ntu$, $i=N-D_{\textrm{max}}+1,...,N$.  As expected, an increase in number of paths with $D_{\textrm{max}}$ results in an increase in number of paths combinations with a maximal differential delay $\tau_{\textrm{up}}(r)$, and, thus, an increase in $P_{\textrm{up}}(r)$. In this scenario, the occurrence probability of a maximal differential delay increases with increasing number of incoming lanes $k$, e.g., $P_{\textrm{up}}(0)\approx6\%$ and $P_{\textrm{up}}(0)=24\%$ in transmission system with $n=k=4$ and $n=k=8$, respectively, and $D_{\textrm{max}}=5$. At the same time, when there was a large number of redundant parallel paths $r$, the occurrence probability of a maximal differential delay $\tau_{\textrm{up}}(r)$ significantly decreases. The simulations match the theory based on Eqs.~\eqref{ProbLongest4} and ~\eqref{ProbLongestR4}.

\par
\begin{figure}[t]
\includegraphics[width=1\columnwidth]{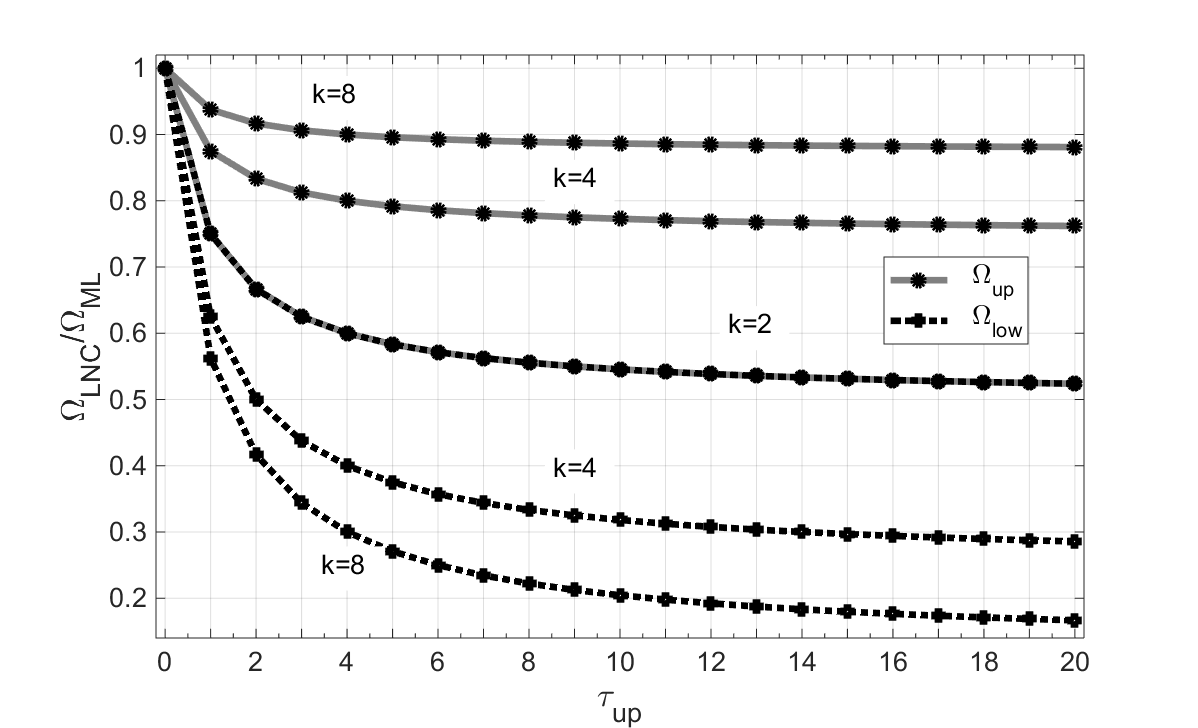}
\caption{The absolute upper and lower bounds of decoding buffer for LNC-OPT and LNC-RND.
}\label{BufferUpLow}\vspace{-0.4cm}
\end{figure}

 We next compare the upper and lower bounds of decoding buffer size with deskew buffer size. Since LNC-OPT and LNC-RND have the same values of absolute upper and lower bounds, we analyze the RLNC method normalized over ML-OPT, see Eq.~\eqref{Multipath}, and use queue size expressions $\Omega_{\textrm{up}}$ and $\Omega_{\text low}$ defined in Eqs.~\eqref{LNCup} and~\eqref{LNClow}, respectively. As shown in Fig.~\ref{BufferUpLow}, the decoding queue size of RLNC based methods decreases with increasing differential delay and can be as small as $18\%$ of ML-OPT buffer size, as in our case study. Per definition, there is no difference between upper and lower bounds, when parallel transmission is realized over two parallel paths only. In case of zero differential delay, deskew and decoding buffers are equal, as per Eqs.\eqref{decoding}, \eqref{decodingExpexact} and \eqref{Multipath}. In presence of differential delay, i.e., $\tau>0$, the absolute upper bound of decoding buffer for RLNC-based parallel systems is always smaller than in the systems without RLNC.

\par
\begin{figure}[t]
\includegraphics[width=1\columnwidth]{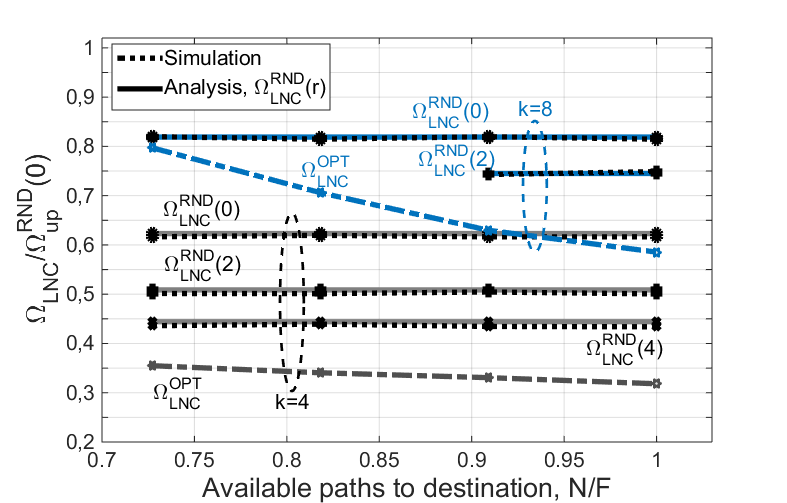}
\caption{Normalized decoding buffer vs number of redundancies and amount of available parallel paths in nsfNet$\_5$
}\label{bufferNet5}\vspace{-0.4cm}
\end{figure}

 Since the upper bound of decoding buffer $\Omega_{\textrm{up}}^{\textrm{OPT}}(0)$ and $\Omega_{\textrm{up}}^{\textrm{RND}}(r)$ (Eq.~\eqref{LNCUp}) generally depend on the pattern of the path delays, i.e., network topology, and are less than or equal to the absolute upper bound $\Omega_{\textrm{up}}$, we next compare the decoding buffer sizes for LNC-OPT and LNC-RND in nsfNet$\_5$, with the buffer size $\Omega_{\textrm{LNC}}$ normalized by the theoretical upper bound $\Omega^{\textrm{RND}}_{\textrm{up}}(0)$, defined by Eq.~\eqref{LNCUp}. Fig.~\ref{bufferNet5} shows that the decoding queue size of LNC-OPT decreases with number of available paths to destination and is about $31\%$ and $58\%$ of the upper bound of buffer size for LNC-RND $\Omega^{\textrm{RND}}_{\textrm{up}}(0)$, for $k=4$ and $k=8$, respectively. The expected value calculated with Eq.\eqref{decoding} is very accurate. This is due to the fact that LNC-OPT establishes a fixed and optimal differential delay pattern (with a minimal differential delay). For LNC-RND, on the other hand, a constant mean buffer size is required independently of the number of paths available and the queue size decreases with increasing redundancy. For instance, the decoding buffer can be reduced from around $63\%$ to $44\%$ of theoretical upper bound $\Omega^{\textrm{RND}}_{\textrm{up}}(0)$ for level of parallelism $k=4$, with $r=4$. The LNC-RND requires a larger queue size than LNC-OPT, however the simulation results for LNC-RND with $k=r=4$ are very close to the optimal solution. In general, the simulation results are very close to theoretical approximation (Eq.~\eqref{decodingExp2}). The decoding queue size was calculated as $59\%$ and $58\%$ of deskew buffer for LNC-RND and LNC-OPT with $n=k=4$, respectively, and did not reach the theoretical upper bounds $\Omega^{\textrm{RND}}_{\textrm{up}}(0)$ and $\Omega_{\textrm{up}}$ (see Fig.~\ref{BufferUpLow}). Moreover, the decoding buffer size $\Omega^{\textrm{RND}}_{\textrm{LNC}}(0)$ is reduced by $41\%$ and $48\%$ for LNC-RND compared to the deskew buffer, with $n=k=4$ and $n=k=8$, respectively, which validates the statement in Eq.~\eqref{RNDbuffercompare}.

\begin{figure}[t]
\includegraphics[width=1\columnwidth]{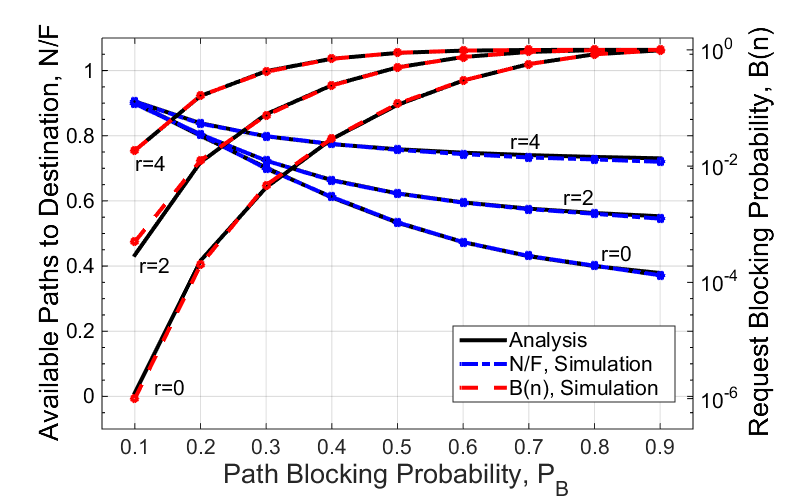}
\caption{Normalized amount of available paths $\bar N$ and blocking request probability $B(n)$ vs. path blocking probability $P_{\textrm{B}}$.
}\label{NvsB4}\vspace{-0.4cm}
\end{figure}

\begin{figure}[t]
\includegraphics[width=1\columnwidth]{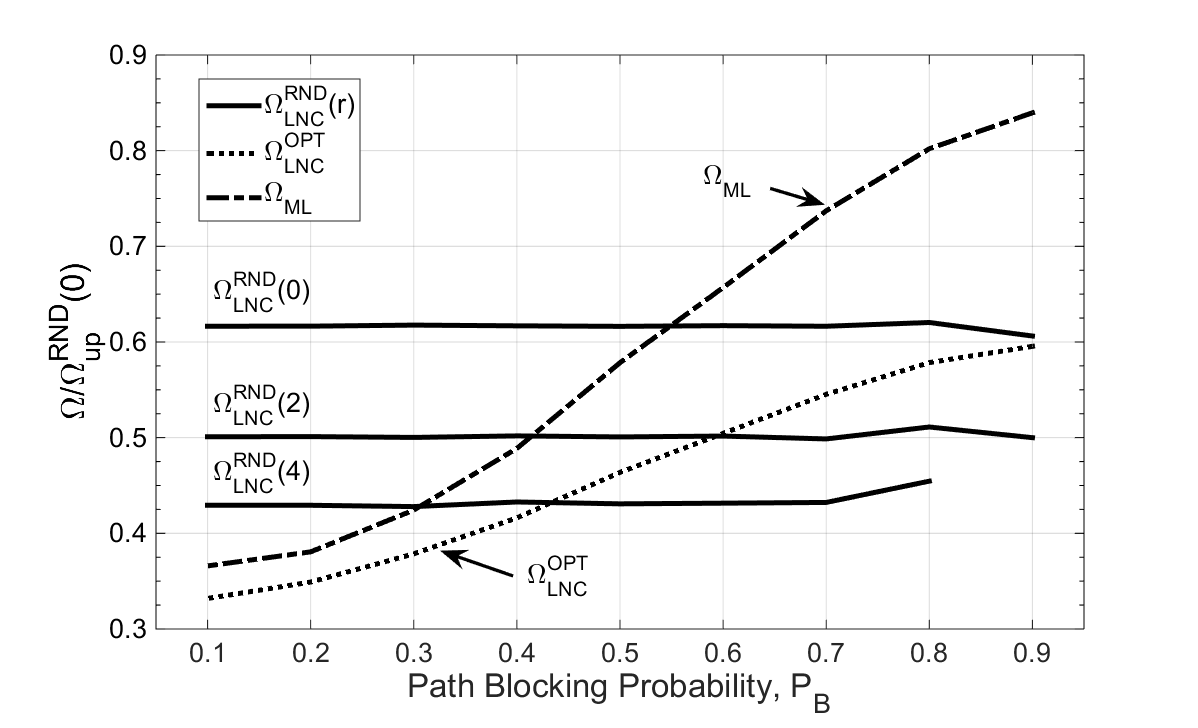}
\caption{Buffer size vs. path blocking probability ($k=4$).
}\label{BuffBlocking}\vspace{-0.4cm}
\end{figure}

\par We next analyze the scenarios where paths can be blocked.  The network studied is nsfNet$\_5$, where each existing path can be available with probability $P_{\textrm{setup}}$, or be blocked with probability $P_{\textrm{B}}$ (Eq.~\eqref{blocking}). In the simulation, this was implemented as ergodic process and followed Bernoulli distribution. In the simulation, the transmission was successful when at least $n$ requested paths were available, while paths optimization was only possible, when network provided $N> n$ paths.

\par Fig.~\ref{NvsB4} shows the mean number of available paths $\bar N=E\{N|N\geq n\}$ (Eq.~\eqref{meanPaths1}), normalized by the number of all existing paths $F$ and the request blocking probability $B(n)$ (Eq.~\eqref{ProbBlockedTrans}) as a function of path blocking probability $P_{\textrm{B}}$. Here, the results for transmission without redundancy, i.e., $r=0$ mean ML-OPT, LNC-OPT and LNC-RND parallel transmission methods over $n=k$ paths. As can be seen in Fig.~\ref{NvsB4}, the mean number of available paths $N$ increases with number of requested parallel paths $n=k+r$. That results, on the other hand, in a very large request blocking probability $B(n)$ calculated with Eq.~\eqref{ProbBlockedTrans}, which is almost $100\%$ in the presented example for path blocking probability $P_{\textrm{B}}$ larger than $30\%$. A reduction in number of redundancy $r$ leads to fewer blockings $B(n)$, sufficiently low for $k=4$ and $r=0$ as long as $P_{\textrm{B}}<30\%$, while all system configurations resulted in request blocking  $B(n)=100\%$ in case of $P_{\textrm{B}}=90\%$.

\par Fig.~\ref{BuffBlocking} illustrates that LNC-RND with redundancy can outperform the methods with path optimizations (LNC-OPT and ML-OPT) in terms of the buffer size, even for a larger path blocking probability. The buffer requirement for LNC-RND is nearly constant for all values of path blocking probability $P_{\textrm{B}}$ and can be reduced up to $43\%$ of upper bound $\Omega^{\textrm{RND}}_{\textrm{up}}(0)$ defined with Eq.~\eqref{LNCUp} by increasing the amount of redundant paths up to $r=4$. In contrast, the buffer requirement in case of LNC-OPT and ML-OPT increases with increasing $P_{\textrm{B}}$ from $33\%$ to $60\%$ and from $37\%$ to $85\%$, respectively, while the LNC-RND without redundancy showed the unchanged buffer size of about $61\%$ of the upper bound (Fig.~\ref{BuffBlocking}).

\section{Conclusion}
{\par This paper presented a performance study of future network systems that exploit parallelism as new network and end-system abstraction, which we defined as the network and end system ability to split the data flow and forward onto multiple interfaces and links for end-to-end transmission. We focused on a cross-layer case study of high-speed Ethernet-over-optical networks, with implementations of random Linear Network Coding (RLNC) as its specially suited example feature. The results showed a great promise of parallel network systems in general and applications of linear network coding in particular: with a proper set of design parameters, {we were able to show analytically that the buffer size at the receiver can be reduced significantly, the cross layer design simplified and routing eliminated; the latter feature especially interesting for networks with complex routing mechanisms like in optical networks. }

\par By deriving the upper and lower bounds as well as an expected value of the differential delay, we showed that a system with RLNC always requires a buffer smaller than in a system without RLNC. We derived analytically the expected values of the differential delay of randomly routed networks with RLNC, and showed that it was independent of path blocking and number of available paths, which is an interesting result. This allows us to make use of more suboptimal paths in the network, whereby it was showed that a larger number of paths decrease the occurrence probability of the maximum possible differential delay, and furthermore reduce the mean value of expected differential delay, and hence the buffer size. 

\par Future work needs to provide a better understanding of networks with heterogeneous level of parallelism between nodes, dynamically changing path blocking, the transmission overhead of parallelization as well as of complexity of buffer implementations. Since coding redundancy in a lossless network was shown to reduce the mean as well as the upper bound of differential delay, the resulting buffer size at the receiver, the possibility of flexible level of parallelism and adaptive (re)coding in the network also appears as a promising avenue for further research. Deriving expressions to regular topologies, such as for networks in optical data centers, is a straightforward extension and application of our analysis. 
}

\section{Appendix}

{
\subsection{ Derivation of Eq.(\ref{ProbTi}) }
 The delays of the $N$ paths $\mathcal P_l \in  G_{\mathcal P}$ are sorted in the ascending order, i.e., $d_{1} \leq ... \le  d_m \le   ... \leq   d_{l}... \leq   d_{N}$. In the same way, 
 a  subset of $ n$ paths  $\mathcal P_{l_m}(\alpha) \in \mathcal M_{n}(\alpha)$ have delay values $d_{l_1}(\alpha) \le ...  \le d_{l_m}(\alpha)  \le ... \le d_{l_{n}}(\alpha)\}$. Due to this ordering, it is obviously that the delay of the $m^{th}$  path $\mathcal P_{l_m}(\alpha) \in \mathcal M_{n}(\alpha)$ can only be mapped to delays from $\vv d$ in the range $d_m\leq d_{l_m}(\alpha)\leq d_{N-n+m}$. Now let us assume, that the $ m^{th}$ path  $\mathcal P_{l_m}(\alpha)$ of an arbitrary subset  $\mathcal M_{n}(\alpha)$ is mapped to the $ l^{th}$ path  $\mathcal P_l \in  G_{\mathcal P}$ with delay $d_l$. Then there are $m-1$ paths with delays  $d_{l_{1}}(\alpha) \le  d_{l_{2}}(\alpha)  \le ...  \le  d_{l_{m-1}}(\alpha)$ whose delays are smaller  or equal to $d_{l_{m}}(\alpha)$, which can be mapped to a set of $l-1$ paths with delays $d_{1} \le  d_{2} \le ...  \le  d_{l-1}$. Since $l \ge m$ there are $C_{l-1,m-1}$ possible combinations for this mapping. 

Similarly, there are $n-m$ paths with delays  $d_{l_{m+1}}(\alpha) \le   ...  \le  d_{l_{n}}(\alpha)$ whose delays are larger or equal   as $d_{l_{m}}(\alpha)$, which can be mapped to a set of $N-l$ paths with delays $d_{l+1}  \le ...  \le  d_{N}$. Since $l \le N-n+m$ there are $C_{N-l,n-m}$ possible combinations for this mapping. Overall, we have in total $C_{l-1,m-1} C_{N-l,n-m}$ combinations with equal probability, where the $l^{th}$ path $\mathcal P_l$ with delay $d_l$ from $\vv d$ is selected as $m^{th}$ path with delay $d_{l_m}(\alpha)$, thus this probability is given by

\par  $p_l(m)=  1/a_{\mathcal M}    \sum_{\alpha} \delta_{\alpha}(l,m) =  C_{l-1,m-1} C_{N-l,n-m} /a_{\mathcal M}$

\subsection{Derivation of Eq.\eqref{newMaxBound} }
The routing over network is implemented over $k$ paths and $r$ redundant paths in parallel. Thus, $n=k+r$ paths from $N$ existing paths can be chosen randomly. For decoding to start, the receiver needs only $k$ data blocks from any those $k$ paths that are shorter than the remaining $r$ paths. Since the vector $\vv d$ is sorted in ascending order as $d_1\leq d_2\leq ...\leq d_{k}$, a maximal path delay that has an impact on decoding start is $d_{k}$, and, thus, the delay $d_{k}$ from $\vv d$ defines a minimal possible value of a maximal path delay. On the other hand, the last $r$ delays from vector $\vv d$ can be any delay $d_l$, ${N-r+1}\leq l\leq {N}$. Since only paths with delays $d_{N-k-r}$, $d_{N-k-r+1}$, ..., $d_{N-k-r+k}=d_{N-r}$ provide data blocks relevant for decoding to start, the path delay $d_{N-r}$ is a maximum delay, which a receiver experiences.
}
\end{document}